\documentclass{aastex}          
\usepackage{spr-astr-addons}    
\usepackage{url}

\newcommand{\wifes}{WiFeS}
\newcommand{\pywifes}{PyWiFeS}

\begin{document}
%
\title{PyWiFeS: A Rapid Data Reduction Pipeline \\ 
for the Wide Field Spectrograph (WiFeS)}

\shorttitle{PyWiFeS}
\shortauthors{Childress, Vogt, Nielsen, \& Sharp}

\author{Michael~J.~Childress\altaffilmark{1,2}} 
\and 
\author{Fr\'ed\'eric~P.~A.~Vogt\altaffilmark{1}}
\and 
\author{Jon~Nielsen\altaffilmark{1}}
\and 
\author{Robert~G.~Sharp\altaffilmark{1}}

\altaffiltext{1}{Research School of Astronomy and Astrophysics, 
Australian National University, 
Canberra, ACT 2611, Australia.}
\altaffiltext{2}{ARC Centre of Excellence for All-Sky Astrophysics (CAASTRO).}

\begin{abstract}
We present \pywifes, a new Python-based data reduction pipeline for the Wide Field Spectrograph (\wifes).  \pywifes\ consists of a series of core data processing routines built on standard scientific Python packages commonly used in astronomical applications. Included in \pywifes\ is an implementation of a new global optical model of the spectrograph which provides wavelengths solutions accurate to $\sim$0.05~\AA\ (RMS) across the entire detector. The core \pywifes\ package is designed to be scriptable to enable batch processing of large quantities of data, and we present a default format for handling of observation metadata and scripting of data reduction.
\end{abstract}


\section{\wifes\ and \pywifes: An Introduction}
The Wide Field Spectrograph \citep[\wifes\ --][]{dopita07, dopita10} is an image-slicing integral field spectrograph built at the Research School of Astronomy and Astrophysics (RSAA) of the Australian National University (ANU). \wifes\ is continuously mounted on the ANU  2.3m telescope at Siding Spring Observatory (SSO) in New South Wales, Australia. The medium wavelength resolution and wide field of view of \wifes\ make it ideal for a multitude of studies such as galaxy kinematics, radial velocity measurements of stars hosting planets \citep{bakos13, bayliss13}, spatially-resolved emission line measurements in galaxies undergoing star-formation \citep{green10} or gas-rich mergers \citep{vogt13}, and the study of narrow emission lines from supernovae interacting with a circum-stellar medium \citep{fraser09ip, inserra12ca}.

The first data reduction pipeline for \wifes\ was developed for use with the NOAO IRAF software and was modeled on the data reduction package for the Near-IR Integral Field Spectrograph \citep[NIFS --][]{mcgregor03}. A need was identified for a rapid data reduction pipeline which could yield fully processed data in nearly real time, thereby adding the capability of real-time classification of astrophysical transients. The \pywifes\ data reduction package was developed to meet this new criterion.

\pywifes\ is designed to make use of existing Python libraries commonly used in astronomical research. The only core requirements are the NumPy, SciPy, and PyFITS packages, with one non-required but strongly recommended dependency on Matplotlib. \pywifes\ is written as a series of data processing routines (functions) that operate on data stored in Flexible Image Transport System (FITS) format and produce output which is also stored in FITS files, without the need for complex new data classes. It is thus intended to be a flexible package which can be run directly via function calls in the Python interpreter, or scripted to perform batch data reduction.

Much effort was expended to make data formats described by abstract variables as much as possible, so that the pipeline is not tailored specifically to \wifes\ except where absolutely necessary. The code is open source and could potentially be adapted for reduction of data from future image-slicing integral field spectrographs, which will play an important role in the next generation of ground-based \citep[e.g., GMTIFS;][]{mcgregor12} and space-based \citep[e.g., WFIRST;][]{spergel13} telescopes.

The focus of this paper is to present the core structure and data processing routines of \pywifes. This software will be continuously maintained by the RSAA and made available to the public via the \pywifes\ Wiki \footnote{http://rsaa.anu.edu.au/pywifes}. The primary Python routines comprising \pywifes\ are briefly described in Section~\ref{sec:routines}, while details on function call syntax and examples are available from the \pywifes\ Wiki. The most complex process in \wifes\ data reduction is derivation of the wavelength solution, which we describe thoroughly in Section~\ref{sec:wavelength_solution}. Finally, the default data reduction script built from the \pywifes\ routines and the preferred \pywifes\ metadata format are described in Section~\ref{sec:top_level}.

\section{\pywifes\ Data Reduction Routines}
\label{sec:routines}
\begin{figure*}[ht!]
\begin{center}
\includegraphics[width=0.95\textwidth]{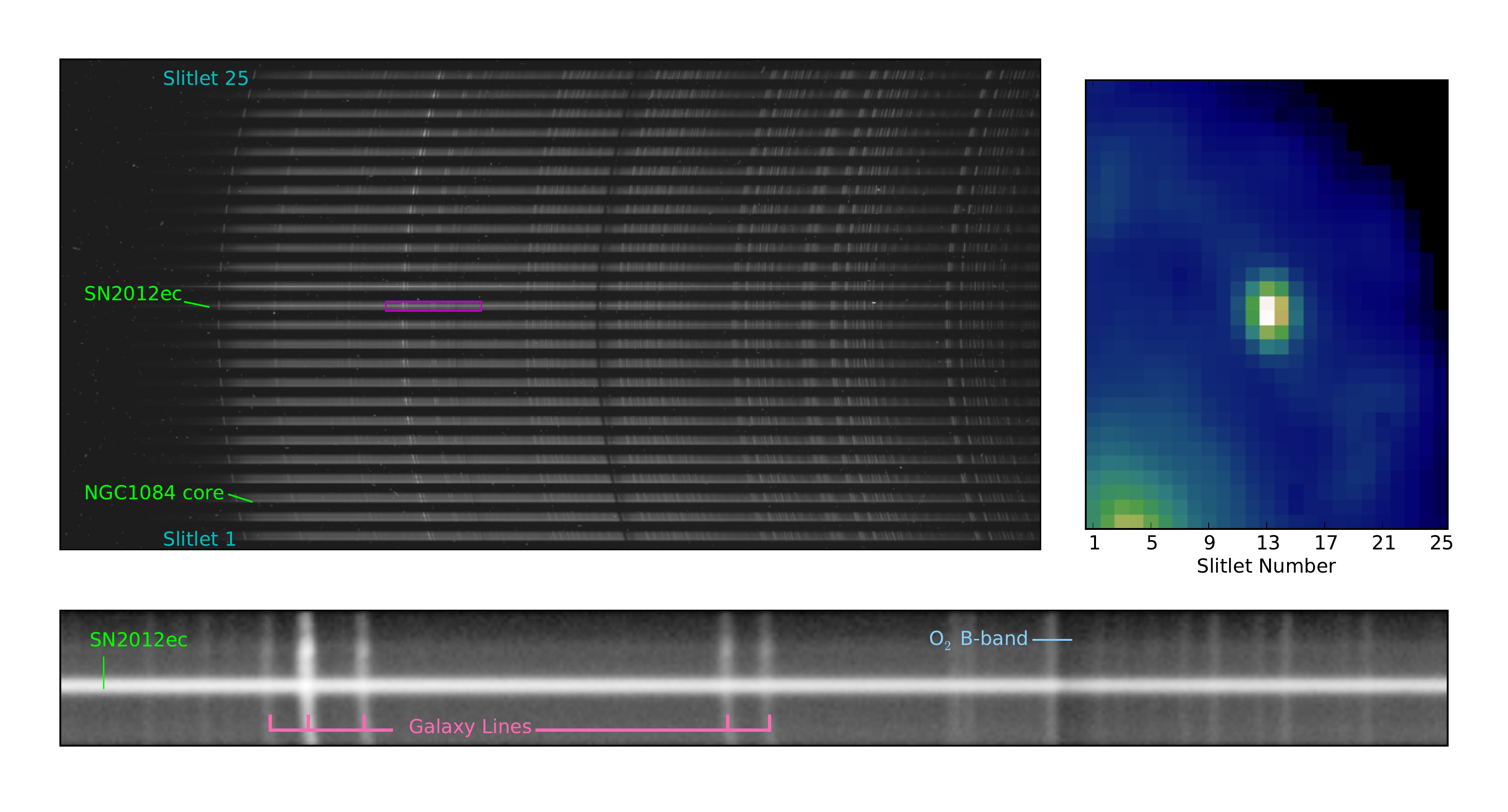}
\end{center}
\caption{Raw data (top left) from a \wifes\ observation of SN~2012ec \citep{maund12ec} in the nearby galaxy NGC~1084. The curved wavelength solution of \wifes\ can be seen from both the night sky lines and the atmospheric O$_2$ A-band (dark band near the center). In the collapse image from the final process data cube (top right) SN~2012ec is visible near the center of slitlet 13, while the core of NGC~1084 is centered near the bottom of slitlet 3. These objects can also be seen in the full raw image. A zoomed-in view (bottom panel) of a region of the raw data (highlighted in purple in the raw data image) shows the SN~2012ec trace, as well as emission lines from NGC~1084 (specifically H$\alpha$ straddled by N~{\sc ii} $\lambda\lambda$6548/6584 along with the S~{\sc ii} $\lambda\lambda$6717/6731 doublet), and the atmospheric O$_2$ B-band.}
\label{fig:raw_data}
\end{figure*}

The \wifes\ instrument has a 25\arcsec$\times$38\arcsec\ field of view (FOV), which the \wifes\ image slicer splits into 25 1\arcsec-wide ``slitlets''. Light from these slitlets passes through a beamsplitter (dichroic) and is sent to the blue and red arms of the spectrograph where it passes through a volume-phased holographic (VPH) grating. This disperses the light into the equivalent of longslit spectra for each slitlet, which are collected on 4k$\times$4k CCDs. An example of raw \wifes\ data from an observation of the supernova SN~2012ec in the nearby galaxy NGC~1084 is shown in Figure~\ref{fig:raw_data}. The profile of each slitlet on the CCD is significantly separated from that of its neighboring slitlet in order to enable nod-and-shuffle observations for optimum sky subtraction \citep[see][for details]{dopita10}.

\pywifes\ image processing routines come in a variety of styles: some operate on raw \wifes\ CCD frames, some operate iteratively on data in each slitlet, and some operate on the global three-dimensional data (commonly referred to as a ``data cube''). We outline the operations employed by the major \pywifes\ processing routines in the sections that follow. Many of the operations require knowledge of certain aspects of the detector characteristics, or default line lists for wavelength solutions, etc. All of this information is stored in a Python dictionary packaged in Python pickle format in the \wifes\ metadata file, which we refer to as needed below. Some routines also require knowledge of the instrument wavelength solution, which is derived using the detailed routine described in Section~\ref{sec:wavelength_solution}, but for this Section we treat it as having already been derived.

\subsection{Image Pre-processing}
\label{sec:bias}
The first step in CCD data reduction is to extract the science pixels from the raw data and convert that data from ADUs to real photon counts (i.e., electrons). This is done by measuring and subtracting the overscan level from the overscan regions of the data, then multiplying the overscan-subtracted ADU values by the gain of the readout electronics. This is done seamlessly for all epochs of \wifes\ data with the \path{subtract_overscan} routine, using detector characteristics stored in the \wifes\ metadata file.

The second step of image pre-processing is to remove any cosmetic defects in the detector. The first generation of detectors (Fairchild CCDs) in \wifes\ were free of any cosmetic defects. In 2013 a second generation of detectors (E2V detectors) were installed in March and May of 2013 for the red and blue detectors, respectively. These detectors provided much improved performance in terms of long-term stability and read noise, but each have a few columns (1 for red, 2 for blue, out of over 4000) of the CCD which are defective in most rows. Correction of these bad pixels is implemented by simply interpolating across the bad columns using the PyWiFeS routine \path{repair_blue_bad_pixels} (and its red counterpart). These bad pixels are also flagged in the data quality extensions when the data is separated into the multi-extension FITS format. Fortunately these bad columns are typically far from any important galaxy emission lines in the red ($\lambda\sim$5550\AA\ for R3000 and $\lambda\sim$5590\AA\ for R7000) and despite potential impact in the blue ($\lambda\sim$4940\AA\ for B3000 and $\lambda\sim$5040\AA\ for B7000), these columns comprise only a few hundred km\,s$^{-1}$ gap in velocity coverage.

\begin{figure}[h!]
\centerline{\includegraphics[width=0.49\textwidth]{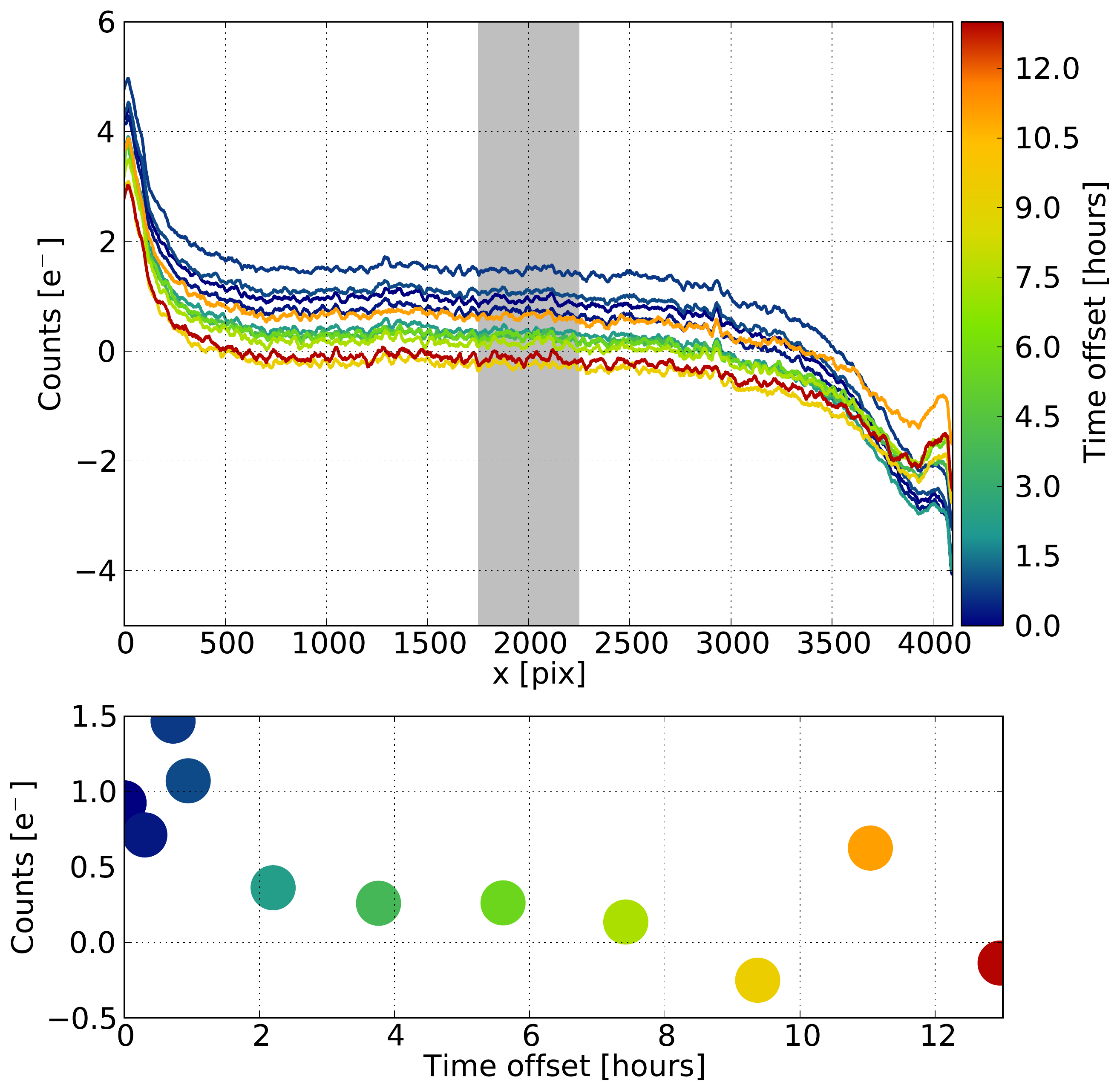}}
\caption{Comparison of 11 blue biases (y-binning of 2 and 1-amplifier read-out mode) acquired over 13 hours  on August 14th, 2012. Top: median bias level of each frame for all x positions. Each curve was smoothed using a Savitzky-Golay routine for clarity \citep[box radius: 51, order: 3, see][]{savitzky64, press07}. Bottom: average of each curve in the top panel, calculated over the grey interval [1750:2250] along x, plotted as a function of the time offset after the first bias was acquired.  }\label{fig:bias_variations}
\end{figure}

Any residual two-dimensional structure in the CCD bias level is removed using bias frames. Typically, this step is performed by median combining several bias frames into a \emph{superbias}. This method is limited by the stability of the detector bias level, which was found to vary both spatially and temporally (on a time scale of minutes to hours at the level of a few e$^{-}$) in the first generation of detectors in \wifes.  We illustrate these variations in Fig.~\ref{fig:bias_variations}, by comparing 11 blue biases observed over a 13 hours interval corresponding to one observing night (including afternoon calibrations). As illustrated in the top panel, the shape of \wifes\ biases is highly non-linear along the x dimension. The mean bias level is subject to variations of ~1 e$^{-}$ within the first hour, and again at the end of the night. The largest shape variations of the mean bias level occur for x positions around 3500 and beyond.

These spatial and temporal variations limit the precision of a standard superbias subtraction method. A solution to this problem, first implemented on \wifes\ data (in a 4-amplifiers read-out mode) by \cite{rich10}, is to fit a multi-dimensional surface to the bias taken closest to any given exposure, and use this \emph{locally reconstructed} bias instead of a global superbias. This bias fitting method was later on officially added to the original IRAF data reduction pipeline for \wifes, and is implemented in \pywifes\ with a slightly different algorithm.

\begin{figure}[h!]
\centerline{\includegraphics[width=0.5\textwidth]{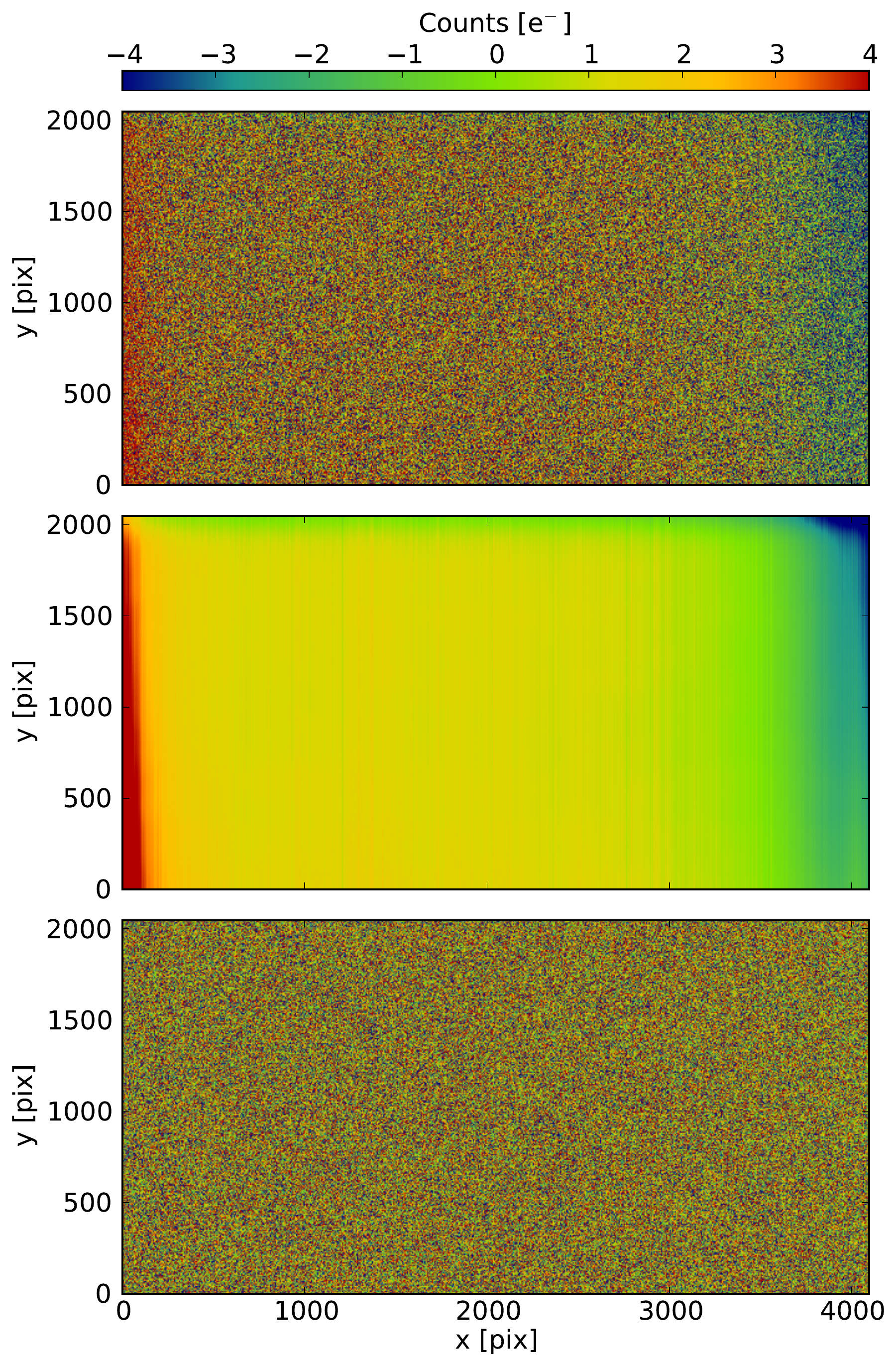}}
\caption{Top: blue, raw superbias, obtained from combining 5 bias frames previously overscan-subtracted individually. Middle : reconstructed, noise-free, superbias. Bottom: raw superbias minus the reconstructed superbias. }\label{fig:bias_sub}
\end{figure}

A typical blue, raw \emph{superbias} (y-binning of 2 and 1-amplifier read-out mode) is shown if Fig.~\ref{fig:bias_sub} (top frame). The underlying structure, although somewhat masked by noise, is clearly visible. Red biases display a similar behavior, and are not illustrated here. Typically, a large number of bias frames are co-added (using the \path{imcombine} routine in \pywifes) to create a raw superbias, from which the bias structure is fitted in a two step process: 

\begin{enumerate}
\item The raw bias is collapsed and median averaged along the y dimension for all x position (see the top panel of Figure~\ref{fig:bias_variations}). This step was introduced first, to correct for the largest spatial variations which occur along the x dimension. 
\item The spatial variations along the y dimension are obtained by performing a 2D smoothing (using a symmetric gaussian kernel of 50 pixels) of the raw superbias minus the 1-dimensional correction obtained in step 1, and subtracted on a row-by-row basis.
\end{enumerate}

In Fig.~\ref{fig:bias_sub}, we show in the middle panel the reconstructed superbias based on the raw superbias in the top panel. In the bottom panel, we show the residual after subtracting the reconstructed bias from the raw bias. As expected, the large scale spatial variations have been removed and the residual is flat at a sub noise-level. 

In the default reduction script (see Section~\ref{sec:script}), this master \emph{reconstructed superbias} is used as the default to perform the bias-subtraction step on any given exposure. It is usually constructed from a series of biases acquired during the afternoon. Individual bias frames acquired by the observer throughout the night can be associated with particular science frames (see Section~\ref{sec:metadata}), and will be used by the reduction script to create \emph{local reconstructed biases} which are then subtracted from the requested science frame. Although these local reconstructed biases rely on individual bias frames, they are in fact noise free, and represent the most accurate way to perform the bias subtractions step with the \wifes\ instrument with first-generation detectors, and best account for the spatial temporal variability of their bias levels.

It is unclear at this stage how the bias levels of the recently installed second-generations detector behave. Once they will have been characterised, \pywifes\ will be updated (if required) to ensure the most appropriate bias subtraction method is being used.

\subsubsection{Separating the Slitlets}
\label{sec:find_slit}
\begin{figure}[h!]
\begin{center}
\includegraphics[width=0.45\textwidth]{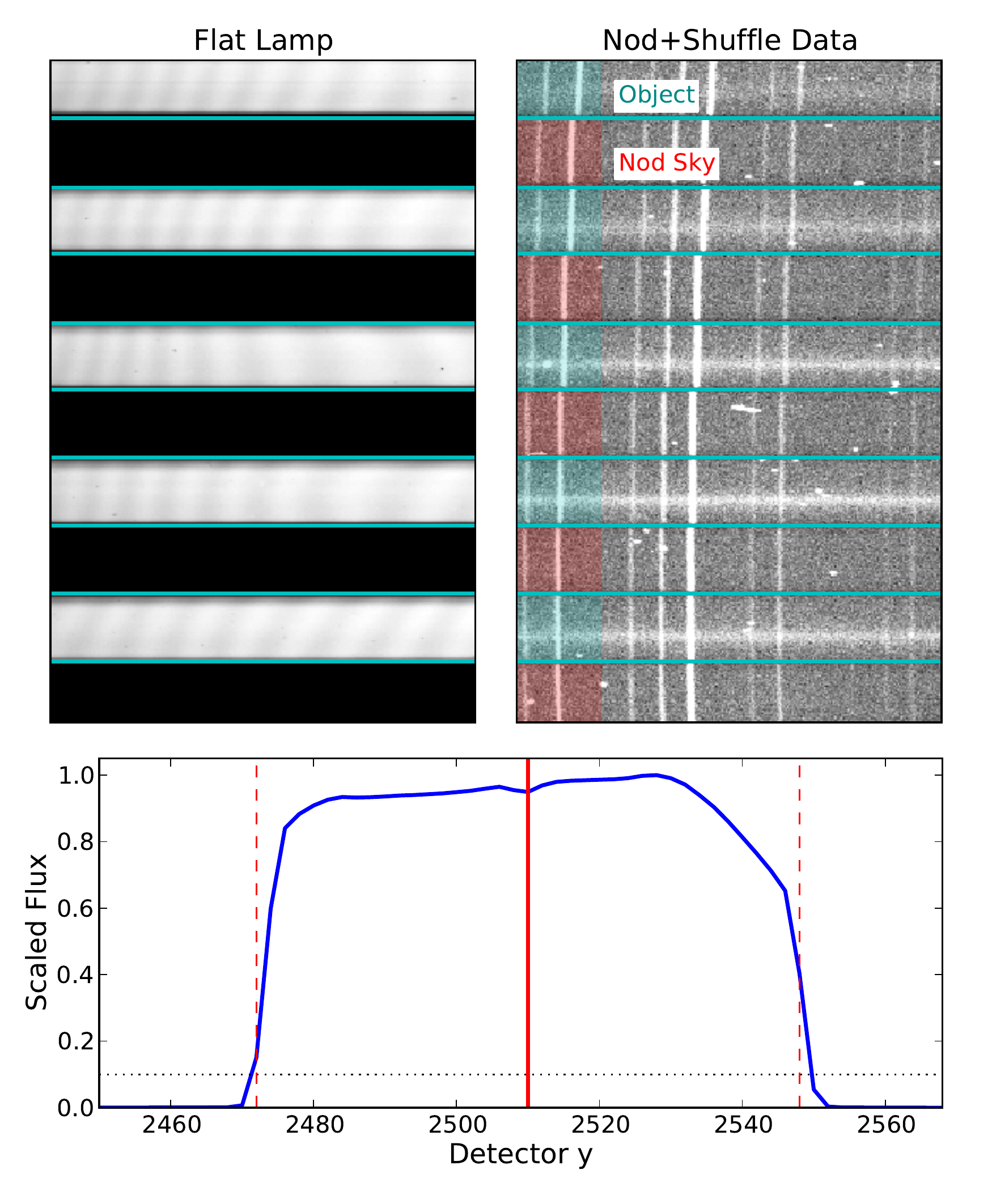}
\end{center}
\caption{Top Left: Flat lamp observation used to measure the slitlet boundaries (cyan lines). Top Right: Nod-and-shuffle (N+S) observation, with object slitlets shaded cyan and nodded sky slitlets shaded red. The \wifes\ cameras were carefully designed to remove distortion in the y direction, allowing for separation of the slitlets by simply selecting the relevant detector rows. Bottom: Fit of the slitlet boundaries (vertical dotted red lines) and center (vertical solid red line) from the flux vs. row profile of the flat lamp observation, using a threshold of 10\% of the maximum flux (horizontal dotted black line).}
\label{fig:slitlet_defs}
\end{figure}

After basic image pre-processing, reduction algorithms for \wifes\ data generally operate on individual slitlets. This requires data for each slitlet to be isolated from the full CCD frame, which is accomplished by first measuring the regions of the CCD occupied by each slitlet. These ``slitlet profiles'' are measured from a flat lamp image by calculating the normalised flux in the flat exposure as a function of the vertical direction on the CCD. The center of the slitlet is measured as the midway point between the two edges of the slitlet, which we define as the highest and lowest rows for which the flat lamp flux exceeds 10\% of the maximum flux. The slitlet is then defined as the 86 rows of the CCD centered on that middle row. This principle is illustrated in Figure~\ref{fig:slitlet_defs}.

Data from the full \wifes\ CCD is split into individual slitlets and saved in a multi-extension FITS (MEF) file with the \path{wifes_slitlet_mef} routine. The first extension of this file contains the original FITS header, as well as some keywords added by the preprocessing routines. The next 25 extensions contain the pre-processed data from each slitlet. The middle 25 extensions are the variance extensions, which when created account for the Poisson noise from photon counting as well as the detector read noise. A final set of 25 extensions serve as ``data quality'' extensions, which are currently only used to flag pixels interpolated over in the cosmic ray rejection step or bad pixels of the second generation detectors repaired in preprocessing. These data quality extensions are built into the data format to allow for complex data quality flag tracking in future incarnations of \pywifes.

For observations in nod-and-shuffle (N+S) mode, the pixels between slitlets are used to store photon counts from nodded sky field observations (see top right panel of Figure~\ref{fig:slitlet_defs}). A variant of the \path{wifes_slitlet_mef} routine designed for N+S observations, \path{wifes_slitlet_mef_ns} saves data from the nodded sky slitlets into a second MEF file specified by the user. Automatic identification of N+S observations and calls to the appropriate MEF saving function are incorporated in the default data reduction script (see Section~\ref{sec:top_level}).

\subsection{Cosmic Ray Rejection}
Cosmic ray (CR) rejection for \wifes\ data is accomplished in \pywifes\ by means of a custom implementation of the Laplacian kernel technique devised by \citet{vandokkum01}. The \pywifes\ CR rejection routine performs additional operations on the data to account for the slanted (sometimes curved) wavelength solution of the instrument, as well as the multiplicity of operating on twenty five slitlets of data.

\begin{figure*}[ht!]
\begin{center}
\includegraphics[width=0.90\textwidth]{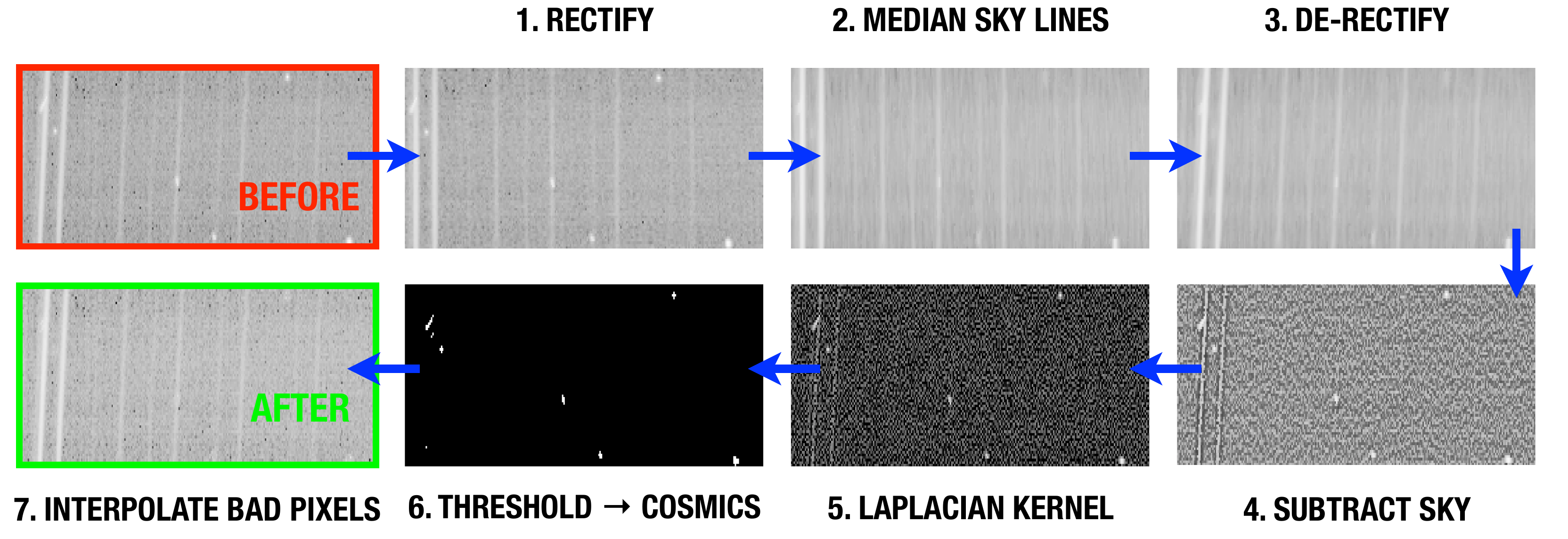}
\end{center}
\caption{Schematic diagram describing the cosmic ray rejection procedure employed for \wifes\ data in \pywifes. The core algorithm is based on \citet{vandokkum01}, but requires iteration over all slitlets (accomplished via parallel processing) and significant modification to account for the slanted (even curved near detector edges) wavelength solution for \wifes.}
\label{fig:cosmic_rays}
\end{figure*}

A schematic representation of the CR rejection steps is shown in Figure~\ref{fig:cosmic_rays}. The first step in identifying CRs is to subtract a smooth model of the sky background. This is complicated for \wifes\ data by the slanted wavelength solution, so the sky model is derived in three steps: (1) resample the data to a rectilinear wavelength grid, (2) smooth with a box-median kernel along the cross-dispersion ($y$) direction, then (3) transform the smoothed sky spectrum back to the original pixel wavelength sampling. This sky model is then subtracted from the original data, and that subtracted data is convolved with a Laplacian kernel as devised by \citet{vandokkum01}. A threshold based on noise statistics of the data and detector read noise is then applied to the convolved data and those pixels above the threshold are identified as being contaminated by CRs. A slightly lower threshold is then applied to neighboring pixels to those identified as CR-contaminated. The data in all CR-contaminated pixels is then replaced with by interpolating from nearby CR-free pixels, and the entire process is repeated for the requested number of iterations (the default is three). The CR rejection procedure is performed for all 25 \wifes\ slitlets, and can take advantage of parallel processing capabilities by simply passing the keyword argument \path{multithread=True} to the CR rejection function.

\subsection{Flat-Fielding}
\subsubsection{Constructing the Response Function}
The throughput of the \wifes\ instrument varies pixel to pixel. At each point, the throughput is a product of losses along the optical path of the instrument and the quantum efficiency of the detector pixel, all of which can vary with position and wavelength. The smooth wavelength-dependent throughput of the instrument (and the atmosphere) is corrected by observing spectrophotometric standard stars, as discussed in detail in Section~\ref{sec:flux_cal}. The spatial variations in the overall throughput are corrected by means of \emph{flat-fielding}, where observations of a spectral source that is both spatially and spectrally smooth is used to directly measure and correct these spatial variations.

\begin{figure*}[ht!]
\begin{center}
\includegraphics[width=0.95\textwidth]{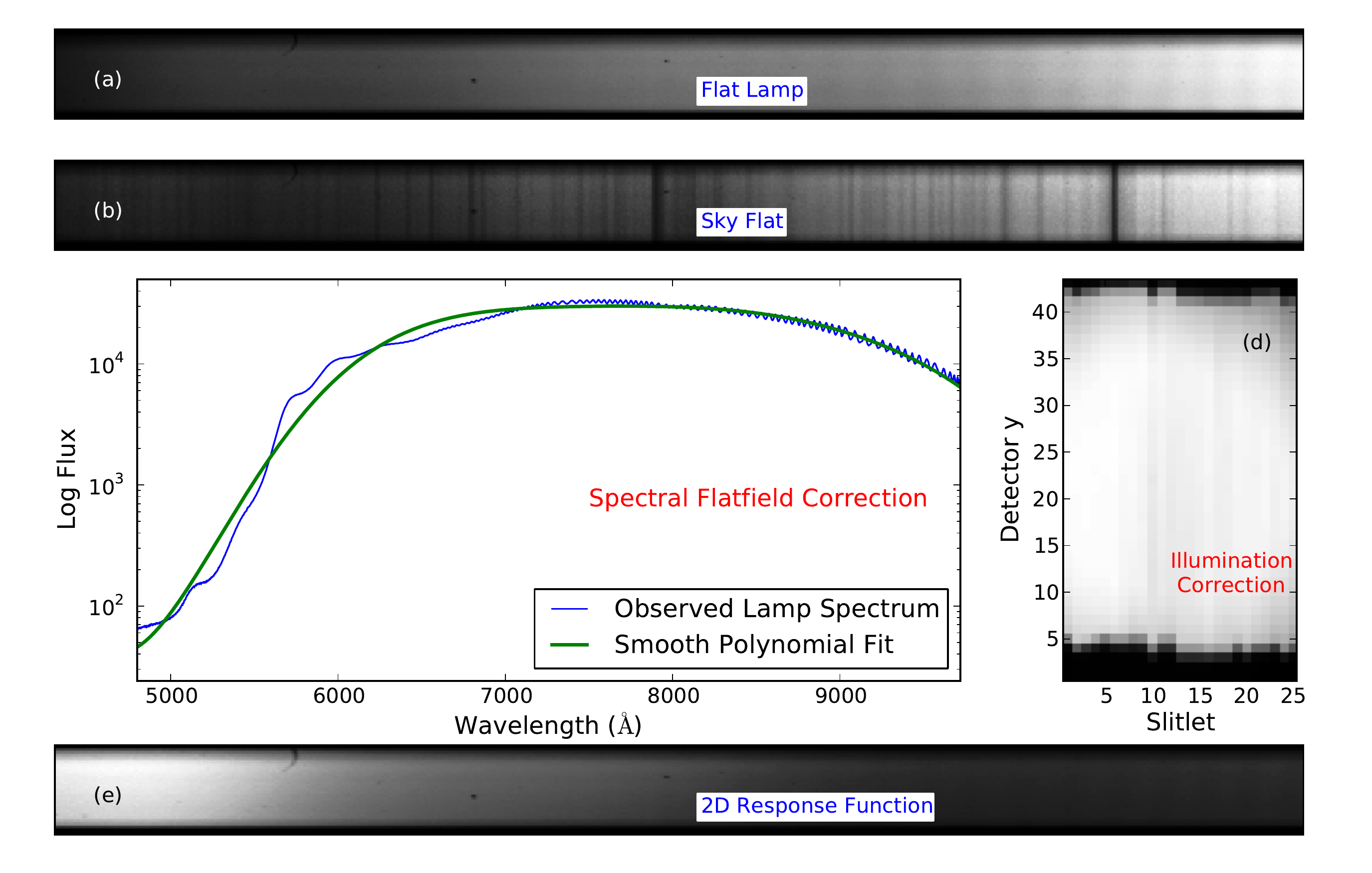}
\end{center}
\caption{(a) Flat lamp data, with ``cold pixels'' and fringing effects clearly visible. (b) Sky flat (twilight) data. (c) Fit of smooth function (green curve) to the flat lamp spectrum (blue curve), which corrects for cold pixels, fringing, and ``dichroic wiggles''. (d) Illumination correction determined from sky flat data (before correction of the vertical offsets between slitlets). (e) Final two-dimensional master response function used to flatfield data.}
\label{fig:flatfield}
\end{figure*}

For \wifes, the smooth spectrum lamp source is part of the instrument's internal calibration unit. Our tests have indicated that the spatial illumination from the internal calibration unit differs slightly from the on-sky optical path. For this reason, a complete flat-fielding solution requires a complex combination of smooth spectrum source data from the internal lamp unit (\emph{lamp flat} observations) and spatially flat on-sky data obtained by observing the ambient glow of the twilight sky (\emph{sky flat} observations). These two types of calibrations provide respectively the spectral flat-field correction and the spatial flat-field (illumination) correction, which are derived with techniques outlines below and illustrated schematically in Figure~\ref{fig:flatfield}.

The spectral flat-field correction is derived from lamp flat observations in the following steps:
\begin{enumerate}
  \item The raw spectrum of the flat lamp is first derived by calculating a median across all rows of the center slitlet. (Note that the dispersion in the center slitlet is acceptably close to vertical to achieve a reliable measure of the lamp spectral shape.)
  \item The smooth shape of the lamp spectrum is derived by fitting a low order polynomial to the lamp spectrum in logarithmic flux versus wavelength space. 
  \item Each row of each slitlet is divided by this smooth spectrum shape, after scaling the smooth spectrum to match the observed flux in the middle half of that row (in practice, we use the middle 2000 columns). The ratio of the observed flux to the smooth lamp spectrum shape gives the spectral response for that row. 
  \item Division by the (scaled) smooth spectrum shape is repeated for all rows of all slitlets to derive the final spectral flat-field correction.
\end{enumerate}

The spatial flat-field correction (often called the ``illumination'' correction) is derived from sky flat observations, taking advantage of the fact that each row of each slitlet corresponds to roughly a single spatial pixel (``spaxel'') on the sky. The illumination correction is derived in the following steps:
\begin{enumerate}
  \item The sky flat is first divided by the spectral response function to correct CCD features or variation in the dichroic throughput. It is expected that the dichroic throughput will vary across the instrument FOV due to the differing angles of incidence and the resulting change in the (angle-dependent) dichroic throughput. This must be corrected before the illumination correction (driven by the other instrument and telescope optics) can be derived.
  \item The baseline spatial flat spectrum is measured from the middle row of the middle slitlet. 
  \item Each row of each slitlet is divided by the baseline spatial flat spectrum (after it has been resampled to the wavelength sampling of that row). 
  \item The illumination correction for each row is calculated as the median value of this ratio (observed spectrum divided by the baseline spatial flat spectrum) in the columns corresponding to a fixed wavelength range, which was chosen to be the wavelength range covered by the middle half (i.e., 2000 columns) of the middle row of the middle slitlet. (Our inspection of the spectrum ratios for each row showed these ratios to be very flat in this central wavelength range, providing a reliable measurement of the throughput for each spaxel.) 
  \item The preceding two steps were repeated for all rows of all slitlets, yielding a spatial flat-field (illumination) correction for all spaxels in the instrument FOV.
\end{enumerate}

The final flat-field response function is the product of the spectral flat-field correction and the spatial flat-field correction. This master response function effectively corrects for three effects: (1) variations in quantum efficiency of particular pixels on the detector (i.e., ``cold'' or ``hot'' pixels), (2) moderate spectral ``wiggles'' in the overall instrument throughput caused by the dichroic beamsplitter (including spatial variation of the wiggles), and (3) non-uniformity in the spatial illumination across the instrument FOV.

\subsubsection{Accounting for Scattered Light}
\begin{figure}[h!]
\centerline{\includegraphics[width=0.5\textwidth]{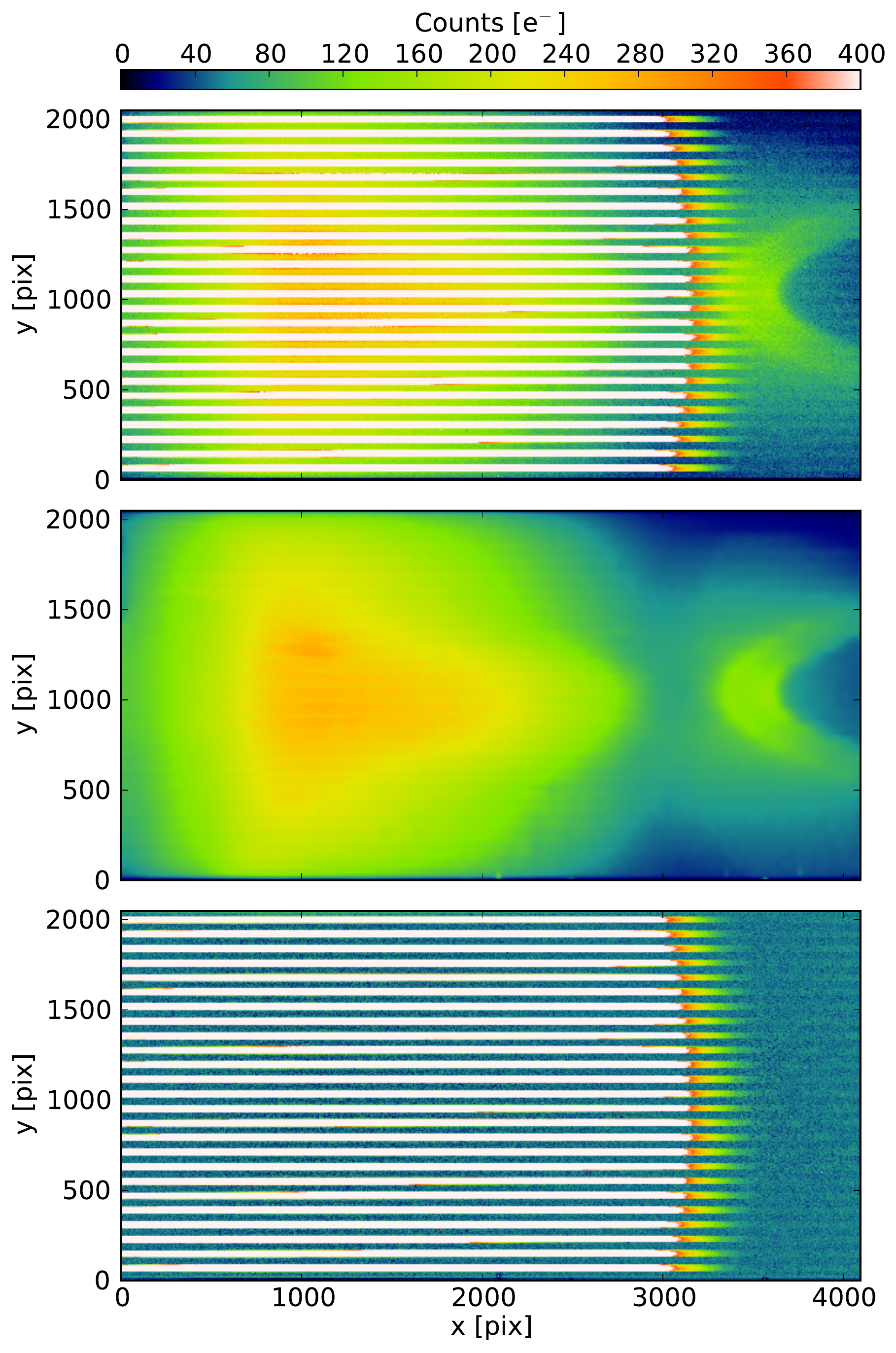}}
\caption{Top: Raw master lamp flat (B3000, y-binning of 2 and 1 amplifier read-out). The internal horseshoe-shaped internal reflection is clearly visible to the right of the image. Middle: reconstructed internal reflection and diffuse glow from the interslit regions. Bottom: cleaned master lamp flat.}\label{fig:flat_sub}
\end{figure}

After the development of the \pywifes\ flatfielding algorithm, examination of the data revealed very subtle irregularities in the spectral flat-field solution which were determined to be caused by scattered light within the instrument. A representative example of a blue master lamp flat is shown in the top panel of Figure~\ref{fig:flat_sub} (red flat-fields are similar). The color scale has been set to reveal the horseshoe-shaped internal reflection to the right of the frame, and also reveals the diffuse light between each of the 25 slitlets visible towards the center of the frame as a red glow. Over most of the spectral range (the x dimension), the diffuse glow only amounts to ~1-2\% of the slitlet fluxes at any given position. To the right of the frame (corresponding to shorter wavelengths), the lamp becomes very faint. The horseshoe reflection largely dominates the signal in this region and strongly affects the flat-fielding of the data below 4500\AA\ if left uncorrected.

\begin{figure}[h!]
\centerline{\includegraphics[width=0.5\textwidth]{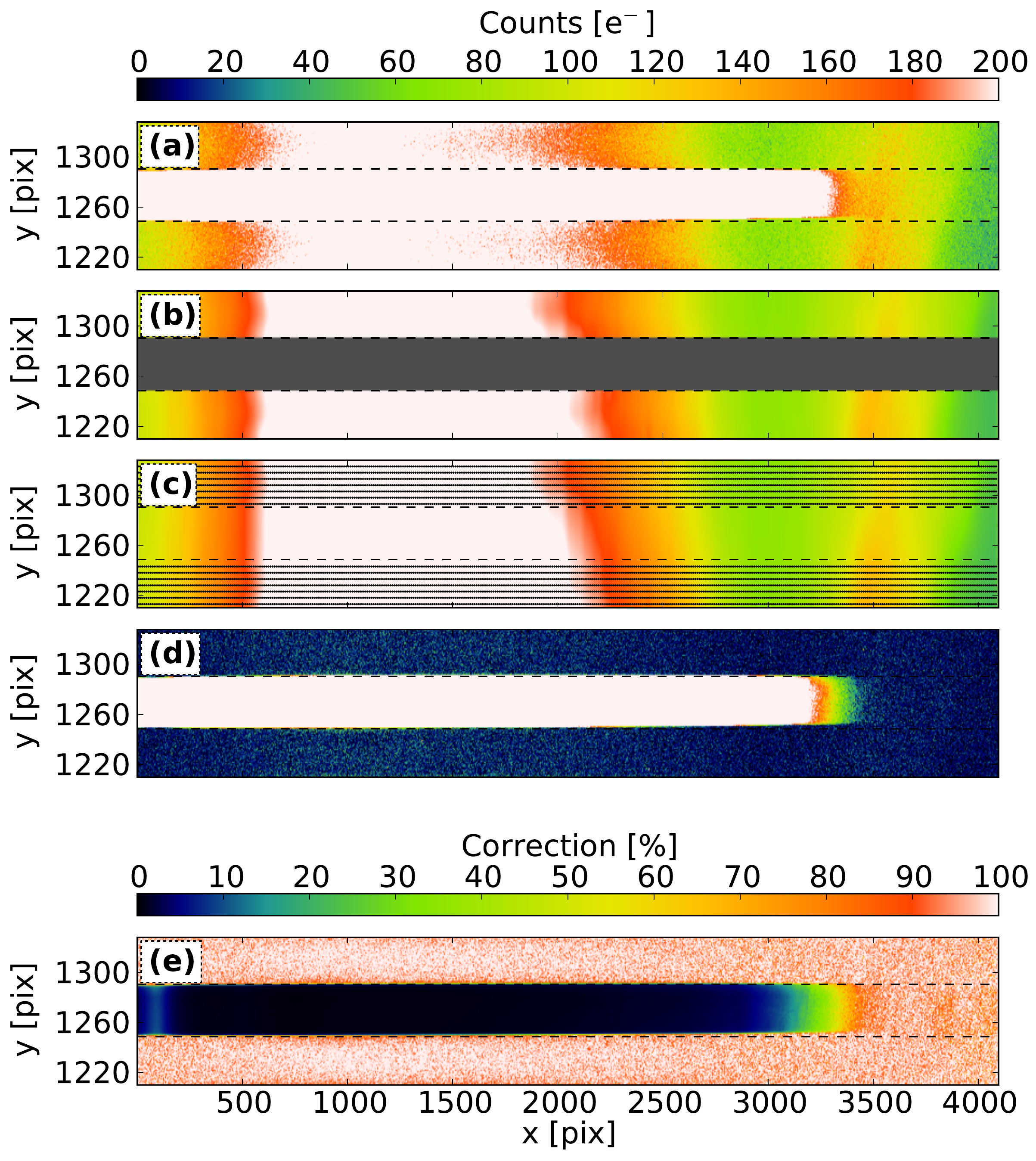}}
\caption{Illustration of the 4 steps implemented in \pywifes\ to remove the internal reflection and the diffuse glow in \wifes\ flats (slitlet number 10). (a) Raw frame (1 slitlet and 2 interslit regions). (b) Smoothed interslit regions only. (c) Smoothed interslit regions, extraction grid points and reconstructed contamination over the slitlet region. (d) Final, corrected slitlet and interlsit regions. (e) Correction intensity in percent of the original data. In each panel, the dashed lines delineate the slitlet/interlsit limits.}\label{fig:deflat_slice}
\end{figure}

To prevent contamination of the flatfield response function by this scattered light, we reconstruct the global internal reflection structures (both the horseshoe and the diffuse glow) from the regions of the detector between slitlets (the ``interslit'' regions), then subtract this correction from the raw master lamp and master sky flats (see Figure~\ref{fig:flat_sub}, middle and bottom panels). The detailed algorithm is a 4 step process, illustrated in Figure~\ref{fig:deflat_slice} for one slitlet, and is as follows :
\begin{itemize}
\item{Step 1:} for all 25 slitlets, extract the slitlet region and the two surrounding interslit regions; panel (a).
\item{Step 2:} from the slitlet definitions obtained previously (see Section~\ref{sec:find_slit}), cut out the slitlet, and smooth the interslit regions with a symmetric gaussian kernel of 10 pixels; panel (b).
\item{Step 3:} extract the interslit count values on a finite grid with a resolution $\Delta$y = 3 and $\Delta$x = 10, and use them as input for a Bivariate Spline fitting routine (\path{RectBivariateSpline} in the \path{scipy.interp} module) to reconstruct the reflections across the slit; panel (c).
\item{Step 4:} subtract the reconstructed contamination from the data; panel (d)
\end{itemize}

In panel (e) of Figure~\ref{fig:deflat_slice}, we show, in percent of the original frame intensity, the amount of correction applied to the slitlet. The correction has only very little effect ($\sim$1-2\%) over most of the slitlet, where the lamp flux is strong, but is very efficient when the lamp becomes less efficient, and the internal horseshoe reflection dominates the signal. 

The Bivariate Spline interpolation routine depends very little on the grid resolution (chosen in Step 3) because the grid points are distributed on either sides of a large gap. As a result, only the grid points closest from the slitlet strongly influence the reconstructed pattern across the slitlet. While it is impossible to exactly separate light traversing the primary optical path from scattered or reflected light, this routine allows removal of both the internal horseshoe reflection and the diffuse interslit glow to within the noise level of the data as illustrated in Figure~\ref{fig:flat_sub} (bottom panel). 

\subsubsection{Correcting for Anomalous ``Sag'' in Data}
A careful inspection of reduced \wifes\ data revealed that even after careful bias correction (see Section~\ref{sec:bias}), a residual gradient in the background of every science frame was present along the x (spectral) direction. This ``sag'' is easily seen in the interslit regions of the bias subtracted frames. In the final reduced frames, this residual signal results in a drop in the spectrum intensity in regions where the flat lamp is faint, presumably due to impact of the sag on the flatfield solution. This effect is especially visible with the B3000 grating below 4200 \AA. 

The \emph{desag} step was implemented to provide an optional way to correct this effect in the data. In practice, it is similar to the flat correction step described above, but is applied to all science frames in the night. Currently we categorise this step as optional in data reduction for two primary reason: (1) in its current form the desag step only works for ``point-and-stare'' observations (and not N+S); and (2) the origin of the residual intensity gradient in the observation background is still unclear. \pywifes\ users who think they require this step to reduce their data are welcome to employ it, but are also strongly encouraged to consult with the developers. Preliminary tests on the new \wifes\ data taken after the detector upgrades suggest that the desag step is no longer required, and should only be considered for \wifes\ users reducing observations taken prior to January 2013.

\subsection{Data Cube Generation}
The format of data generated by integral field spectroscopy is flux values as a function of wavelength at each spatial location on the sky. This sampling of the flux is commonly referred to as a ``data cube'' as it samples flux in three dimensions. This data format requires the three coordinates $(x,y,\lambda)$ be derived for each pixel on the detector. While the $x$ coordinate is simply the slitlet number, determining a consistent $y$ coordinate for all slitlets requires a measurement of the spatial zeropoint in each slitlet. This is accomplished using a ``wire'' frame observation, which we describe in detail in Section~\ref{sec:wire}.

The $(x,y)$ coordinate pairs calculated within the instrument field of view do not correspond to the same physical coordinates on the sky $(\alpha,\delta)$ at all wavelengths. This is due to the influence of atmospheric differential refraction \citep[ADR --][]{filippenko82}, where the wavelength-dependent index of refraction of the atmosphere deflects light of different wavelengths by subtly different values. Calculation of this effect is a solved astrophysical problem, and we describe the implementation in \pywifes\ in Section~\ref{sec:adr}.

For most applications, the irregular spatial and wavelength sampling realised by the actual \wifes\ detector pixels is too complex to easily extract physical quantities over the full instrument FOV. Instead the observed fluxes are resampled onto a regular (rectilinear) grid of $(x,y,\lambda)$ for the final data cube, and this step for \pywifes\ is outlined below in Section~\ref{sec:cube_gen}.

\subsubsection{Spatial Zeropoint Derivation}
\label{sec:wire}
\begin{figure*}[ht!]
\begin{center}
\includegraphics[width=0.90\textwidth]{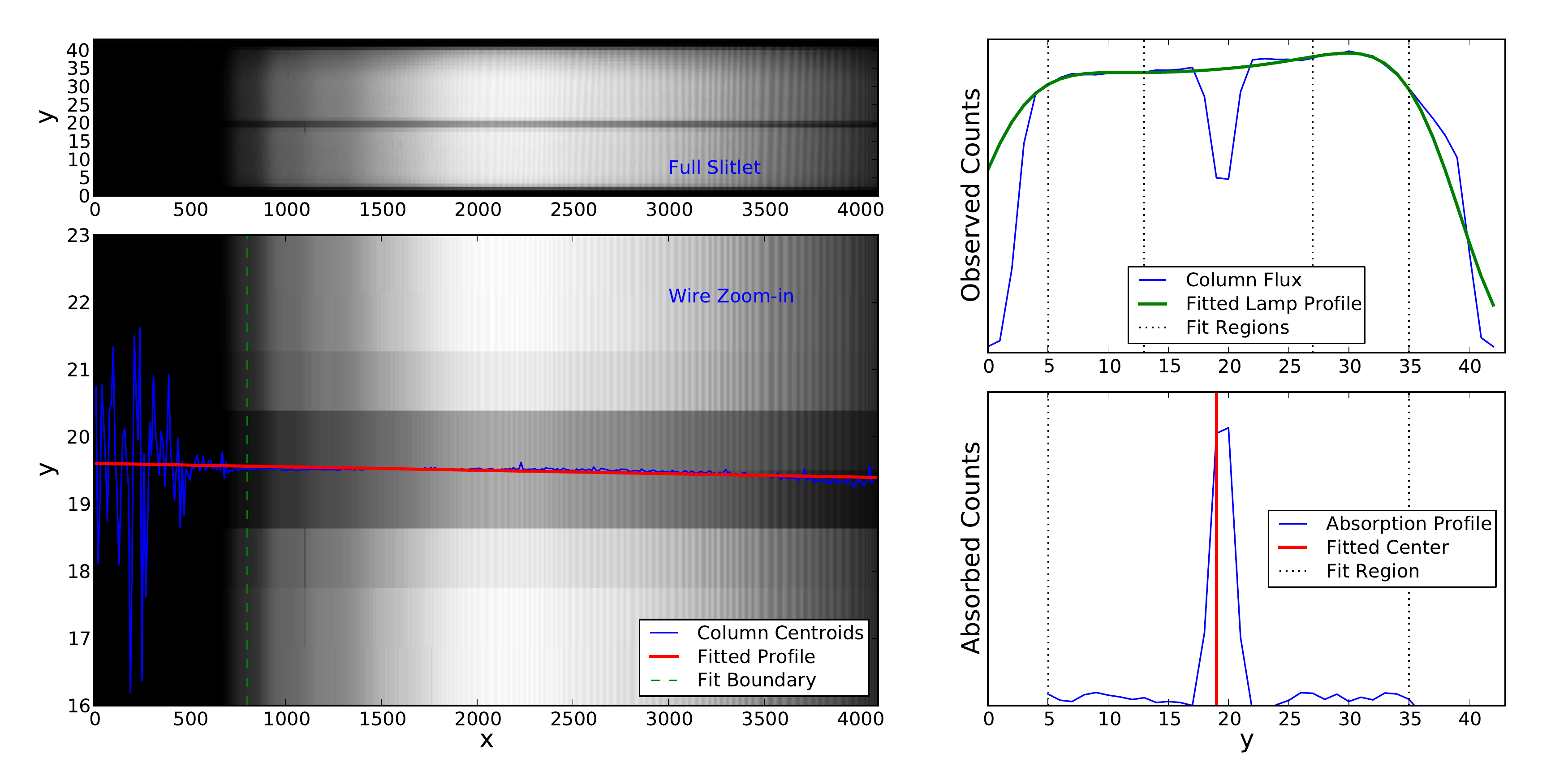}
\end{center}
\caption{Top left: Full slitlet from a wire frame observation. Top right: Column slice from a single slitlet (blue curve). The unabsorbed lamp profile is fitted as a smooth polynomial (green curve) from the regions where the lamp flux is high (dotted vertical lines). Bottom right: Wire absorption profile (blue), with fitted wire centroid (vertical red line). Bottom left: Zoom in of the wire absorption region, with fitted centroids for each column (blue curve), and fitted linear wire profile (red line) from bounded fit region (vertical dashed green line).}
\label{fig:wire}
\end{figure*}

For each column of each slitlet of \wifes\ data, the true vertical coordinate $y$ is determined by measuring the zeropoint with a ``wire frame'' observation. An occulting wire coronograph that is flat in the (true) $y$ direction is placed across the instrument FOV, and the flat lamp is shone onto the detector so that the wire's shadow appears on the detector in each slitlet. The position of the wire is fitted from this shadow using the \pywifes\ \path{derive_wire_solution} routine, whose procedure is outlined graphically in Figure~\ref{fig:wire}.

The wire shadow in each column is isolated by subtracting the observed flux from a smooth continuous model of the flat lamp flux, which is fitted as a low order (default is 4th order) polynomial in detector $y$ over two regions where the lamp flux is high. The wire position is then measured as the centroid of the absorption spectrum in the region bounded by the high lamp flux regions. Once the wire centroid has been measured for each column of the detector, the wire profile is fitted as a simple linear function of detector column, with the fit restricted to columns where the lamp flux is not diminished by the dichroic. The wire solution for all columns of all slitlets is then saved in a FITS file used as input in the data cube generation routine.

\subsubsection{Correction for Atmospheric Differential Refraction}
\label{sec:adr}
\begin{figure*}[ht!]
\begin{center}
\includegraphics[width=0.95\textwidth]{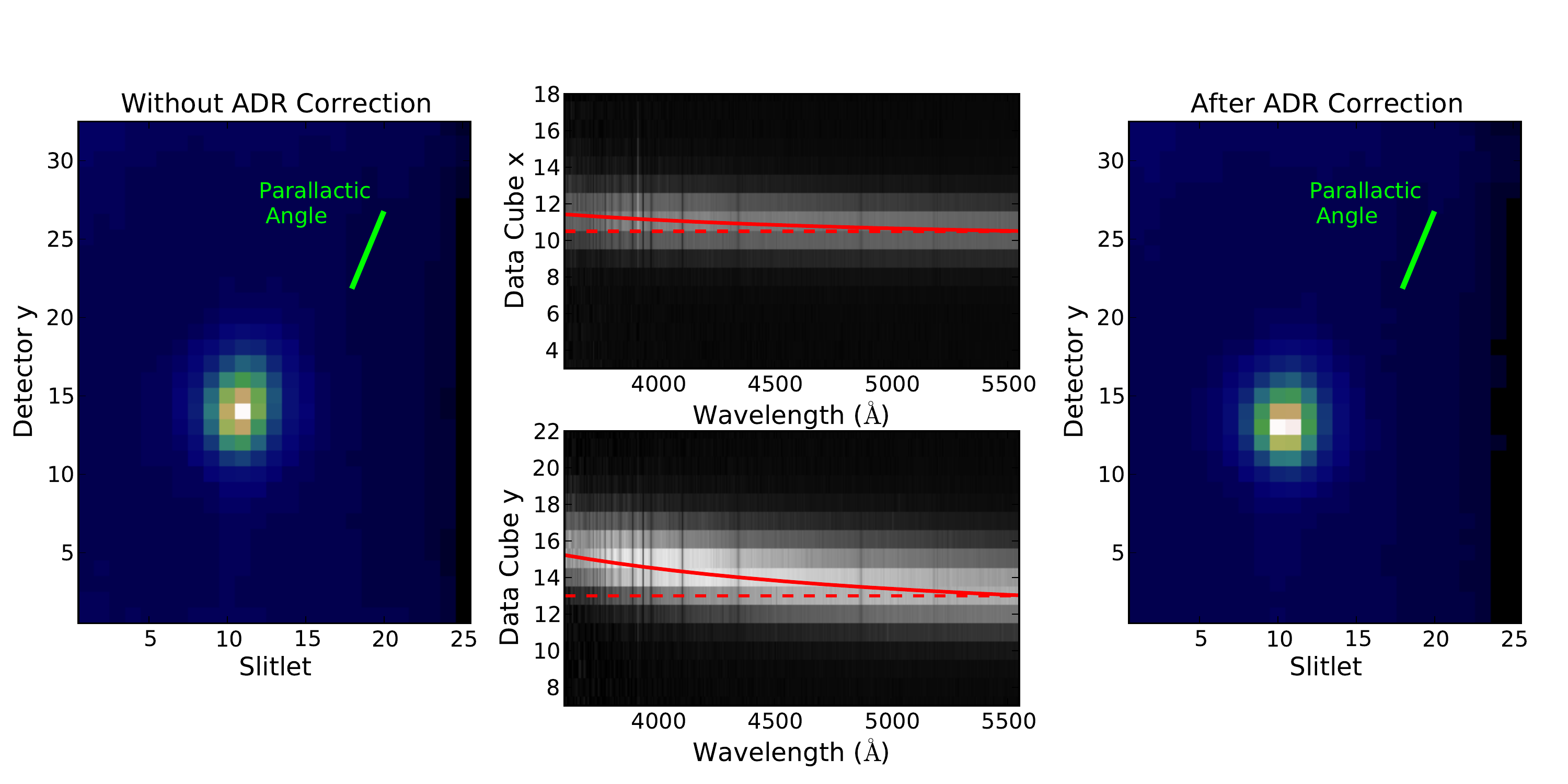}
\end{center}
\caption{Left: Processed \wifes\ data cube of a high airmass ($secz=1.85$) observation of the standard star LTT2415, with no correction applied for ADR. Elongation along the parallactic angle (shown in green) is clearly evident. Middle panels: Collapsed stack of flux vs. $x$ and $y$ spaxel coordinates for the uncorrected data cube. The star trace clearly follows the predictions (solid red curves) of the equations of \citet{filippenko82}, compared to zero deflection (dashed lines). Right: The same \wifes\ data cube but processed with ADR correction. The previous elongation along the parallactic angle has been remedied.}
\label{fig:adr}
\end{figure*}

Atmospheric differential refraction \citep[ADR --][]{filippenko82} causes an excess deflection of light of bluer wavelengths. This causes the apparent position of an object to shift as a function of wavelength, and must be corrected for in integral field spectroscopy data. In \pywifes, this is corrected at the data cube generation step by applying a Python implementation of the ADR equations of \citet{filippenko82}. In Figure~\ref{fig:adr}, we show a high airmass ($secz=1.85$) \wifes\ observation of a standard star (LTT2415). The trace of the star shows excellent agreement (as expected) with the predicted deflection. These deflection curves are calculated directly from the observation details stored in the \wifes\ data FITS header, and applied in the data cube generation step.

\subsubsection{Coordinate Rectification}
\label{sec:cube_gen}
Once spectral and spatial coordinates have been determined for each pixel of each slitlet, the data must then be resampled onto a rectilinear $(x,y,\lambda)$ coordinate system. This is accomplished by performing a two-dimensional resampling of the data for each slitlet. The spatial zeropoint (wire profile) and vertical ADR deflection are used to assign a true $y$ value to each pixel, and the wavelength solution defines the $\lambda$ value for each pixel. The final $(y,\lambda)$ grid for each slitlet is defined by the central 38\arcsec\ of the slitlet, and a wavelength array which by default spans the common wavelength coverage of all slitlets with a pixel scale equal to the mean pixel scale of all the data. Data are resampled from the observed $(y,\lambda)$ coordinates for all pixels to the desired $(y,\lambda)$ grid using the \path{griddata} routine in the \path{scipy.interpolate} Python package.

After this first resampling, correction of ADR deflection {\em across} slitlets requires a second resampling of the data (currently there is no efficient three-dimensional resampling routine in Python, understandably due to the challenging computational requirements). This second resampling is accomplished via a simple one-dimensional resampling along all $x$ values for a fixed $(y,\lambda)$ position. By definition, flux beyond one edge of the detector will be required to properly resample the true flux along that edge. Since there is no way to produce this absent information, the data cube generation routine by default fills this unsampled edge with a copy of the observed edge data, meaning this edge will also have slightly erroneous flux. For this reason, it is still valuable for observers to align their observations along with parallactic angle to avoid this edge resampling problem.

\subsection{Flux Calibration}
\label{sec:flux_cal}
Once \wifes\ data have been corrected for pixel-wise variations of detector efficiency and spatial variations of the instrument throughput, the wavelength dependence of overall throughput of the instrument must still be corrected. This is achieved through observation of astronomical sources with known spectral energy distributions, referred to as spectrophotometric standard stars \citep[see][for a thorough review]{bessell05}. This process of measuring the absolute sensitivity of the instrument and correcting observational data to this absolute scale is known as flux calibration, and is achieved in \pywifes\ using routines defined in the \path{wifes_calib} sub-module. Flux calibration is typically accomplished in three steps:
\begin{enumerate}
  \item Standard star spectra are extracted from data cubes using the \path{extract_wifes_stdstar} routine.
  \item The instrument sensitivity function is derived by comparing observed standard star spectra to their reference spectra using the routine \path{derive_wifes_calibration}.
  \item The flux calibration solution is applied to uncalibrated \wifes\ data cubes using the \path{calibrate_wifes_cube} routine.
\end{enumerate}

\begin{figure*}[ht!]
\begin{center}
\includegraphics[width=0.95\textwidth]{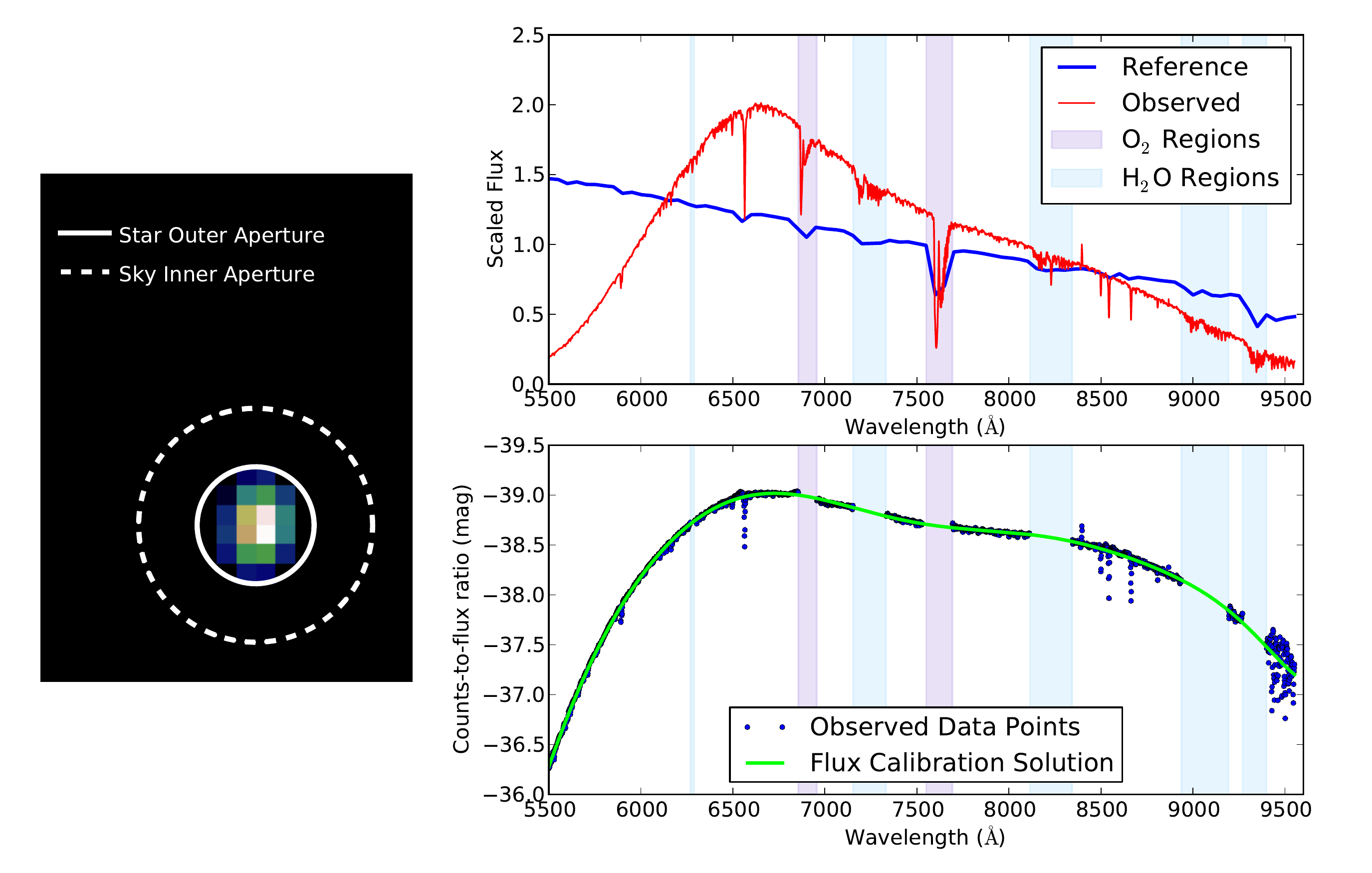}
\end{center}
\caption{Left: Collapsed data cube image for a standard star observation \citep[LTT7379 -- from][]{oke90}, showing the inner boundary of the sky region (dashed line) and the star extraction aperture (solid line). Right Top: Observed flux (red curve -- in photon counts, scaled by its mean flux) compared to the reference star flux (blue curve -- in $F_\lambda$, again scaled by its mean flux). Right Bottom: Flux calibration solution (solid green curve) derived from flux-to-counts ratio (in magnitudes) observed in data (blue points). In the right two panels, telluric regions due to O$_2$ and H$_2$O are shown as shaded purple and blue regions, respectively.}
\label{fig:flux_cal}
\end{figure*}

Standard star spectra are extracted using a simple aperture extraction technique employed by the \path{extract_wifes_stdstar} routine. This is illustrated schematically in the left panel of Figure~\ref{fig:flux_cal}, where the sky background spectrum is estimated as the mean spectrum of spaxels outside some fiducial radius. After this sky spectrum is subtracted from each spaxel, the flux within some object aperture is summed to obtain the final object spectrum. The center of the extraction aperture is determined by fitting the flux centroid in the $(x,y)$ plane, which is generally a robust measure of the star's location. The extraction and sky radii are set by default to 5\arcsec\ and 10\arcsec, respectively, though these values (and the extraction center) can be modified by passing the desired values as keyword arguments to the extraction routine.

After the standard star spectra have been extracted, the instrument sensitivity function can be derived by comparing these to their reference spectra. In \pywifes\ this is accomplished in a (mostly) automated fashion using the \path{derive_wifes_calibration} routine. This function takes a list of standard star cubes as input (or optionally a list of extracted spectra) and extracts all spectra according to the algorithm described above.

The \path{derive_wifes_calibration} routine then attempts to identify the name of the standard star from the 'OBJECT' header field of the data cube file (alternatively the user can pass a list of object names to this routine). These names are compared to a list defined in the \path{wifes_calib.py} file, comprising a Python dictionary containing the reference spectrum file name for each standard star. Future planned upgrades to \pywifes\ include the ability to cross-reference standard star lists by the object coordinates. Once the reference star name has been identified, its reference spectrum is loaded and the ratio of observed flux to reference flux is calculated for each star in the input list. This ratio is then corrected for the smooth atmospheric extinction using the default Siding Spring Observatory extinction curve measured by \citet{bessell99}.

The final flux calibration solution is fitted from the 'counts-to-flux' ratio values (i.e., the sensitivity curves) obtained for all stars in the list passed to \path{derive_wifes_calibration}. Points falling in regions of strong telluric absorption are not included in the fit. The final sensitivity curve can be calculated with a single star or with a list of several stars. The sensitivity curves from all stars can be normalised to one another using the 'norm\_stars' keyword, which normalises all stars based on the middle 200 pixels, scaled to the maximum sensitivity curve.  This allows a higher signal-to-noise measurement of the instrument sensitivity while not being hindered by grey extinction variations throughout the night.

Once the flux calibration solution has been derived, it can be applied to uncalibrated \wifes\ data cubes using the \path{calibrate_wifes_cube} routine. This routine interpolates the derives flux calibration solution to the wavelength values of the input data cube, and multiplies the observed flux (in photon counts) to convert it to the true flux (in $F_\lambda$). This routine also applies the extinction correction, again based on the \citet{bessell99} canonical Siding Spring extinction curve.

We note that the current flux calibration routine in \pywifes\ does not allow for a fit of residual extinction (i.e., deviation from the input extinction curve), though that capability is planned to be included in future upgrades to \pywifes. However, the star spectrum extraction routine is capable of saving the extracted spectrum in a format readable by the IRAF \path{onedspec} package, and the flux calibration routine can accept IRAF-style flux calibration solutions and extinction curves. Thus, users who require stringent flux calibration and extinction corrections can work with existing IRAF routines by employing the appropriate formatting keywords when calling the \pywifes\ extraction and flux calibration functions.

\subsection{Correction of Telluric Absorption}
The broad smooth absorption of light by the Earth's atmosphere is corrected using the atmospheric extinction curve in the flux calibration step in \pywifes. However, narrow structured absorption features caused by molecular species in the atmosphere (primarily molecular oxygen and water) persist in all object spectra after this step. These telluric features are most efficiently corrected by observing sources with smooth spectra at similar airmass and time of the night as a science target \citep[e.g., using the smooth star division technique of][]{bessell99}. Alternatively, measuring telluric absorption in a number of smooth spectrum objects over a range of airmass can provide an acceptable estimate of the average telluric absorption behavior for a given night.

\begin{figure}[h!]
\begin{center}
\includegraphics[width=0.45\textwidth]{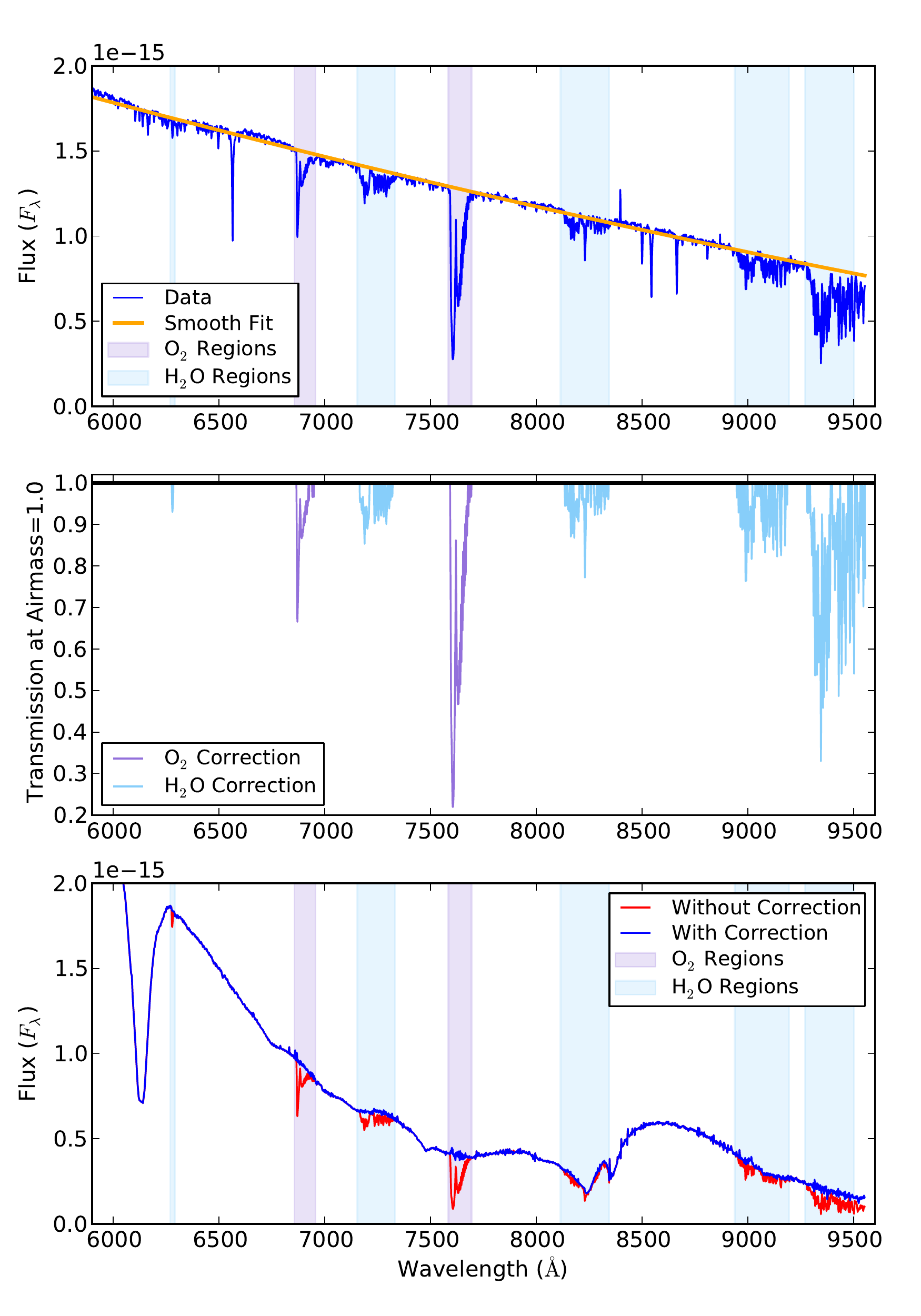}
\end{center}
\caption{Top: Flux-calibrated standard star spectrum, along with low-order polynomial fit of the smooth continuum fitted excluding regions of strong telluric absorption. Middle: Telluric correction functions for O$_2$ and H$_2$O, scaled to airmass $secz=1.0$ (see text). Bottom: Example of a smooth spectrum source \citep[the type Ia supernova SN~2012fr][]{childress12fr} before (red) and after (blue) correction of telluric absorption features.}
\label{fig:telluric}
\end{figure}

In \pywifes, telluric features are measured using the routine \path{derive_wifes_telluric} and corrected using the routine \path{apply_wifes_telluric}. As with the flux calibration routine, the \path{derive_wifes_telluric} routine takes a list of star cubes (or extracted spectra) as input. Regions of the object spectra not affected by telluric absorption are fitted with a low order polynomial, as demonstrated in Figure~\ref{fig:telluric}. The ratio of observed flux to the predicted smooth flux defines the telluric absorption for each spectrum.

Because telluric absorption strength depends on airmass, the absorption from each each spectrum in the list is corrected to airmass $secz=1.0$ and the mean absorption across all objects defines the final telluric absorption correction functions. The airmass dependence is a product of the saturation level of the absorption, which often differs between O$_2$ (which is more saturated) and H$_2$O \citep[see, e.g.,][]{buton13}. The logarithmic power law index for these two species can be set by the user as keyword arguments to the telluric fitting routine, and are stored along with the respective correction curves in a Python pickle (.pkl) file. From an observing night in 2013 we measured these indices to be 0.40 and 0.72 for O$_2$ and H$_2$O, respectively, which are both intermediate between the optically thick (saturated) value of 0 and the optically thin limit of 1. A more detailed study of the airmass dependence of telluric absorption features is planned for future work, but can be easily modified in practice using the keywords built into the routine.

Finally, once the telluric correction functions for O$_2$ and H$_2$O have been determined, they can be applied directly to data (with the appropriate airmass dependence) by calling the \path{apply_wifes_telluric} routine.

\section{\wifes\ Wavelength Solution}
\label{sec:wavelength_solution}
The greatest challenge in the reduction of \wifes\ data is to derive the wavelength solution for the entire instrument from an arc lamp observation. In typical longslit spectroscopy reductions, the wavelength solution is derived from arc lamp data via an interactive procedure where the user finds emission lines by eye and identifies the associated reference wavelength for those lines. The fitted positions of the arc lamp emission lines and their true wavelengths are then used to extrapolate the wavelength value associated with each pixel across the entire detector. Performing line identification manually for multiplexed spectroscopic data such as that obtained with IFUs is not only time consuming, but also particularly prone to human error due to its repetitive nature.

For \wifes\ data, our goal was to develop robust tools for accomplishing the same steps but without requiring the user to identify emission line positions and reference wavelengths by hand. Furthermore, we sought a physical model for the \wifes\ wavelength solution which incorporates the correct analytical parametrization of how the wavelength solution varies across the detector within each slitlet, and how the wavelength solutions of all slitlets are related by the optics of the instrument.

In this Section we outline the algorithms employed in \pywifes\ to find emission lines in arc lamp data and identify their true reference wavelengths (Section~\ref{sec:line_finding}), and the implementation of the \wifes\ optical model in \pywifes\ (Section~\ref{sec:optical_model}). The details of the instrument optical model are beyond the intended focus of this paper, so we defer such details to future work. Instead we focus on software implementation of the model in its application to real data.

\subsection{Finding and Identifying Emission Lines}
\label{sec:line_finding}
Discovering and fitting emission lines automatically from a large volume of data without user interaction requires a robust set of algorithms. For each row of each slitlet, we first identify all pixels which potentially have emission line flux in them with the following technique. We measure the mean and RMS of the background flux level (i.e., from pixels without emission line flux) via an iterative outlier rejection technique. Similarly, we measure the mean and RMS {\em derivative} of flux along the dispersion axis. Pixels must then pass a series of cuts to be identified as a potential emission line center: the flux value must be below the saturation level and above a certain flux threshold, the flux in the adjacent pixels must also exceed the flux threshold, and the derivative of adjacent pixels must exceed a threshold set from the flux derivative statistics (by default thresholds are the mean plus $3\sigma$ for flux and flux derivative as determined from the background statistics). Any adjacent sets of pixels which pass these cuts are grouped together and the potential line center is identified as the middle pixel value of the group.

\begin{figure}[ht!]
\begin{center}
\includegraphics[width=0.45\textwidth]{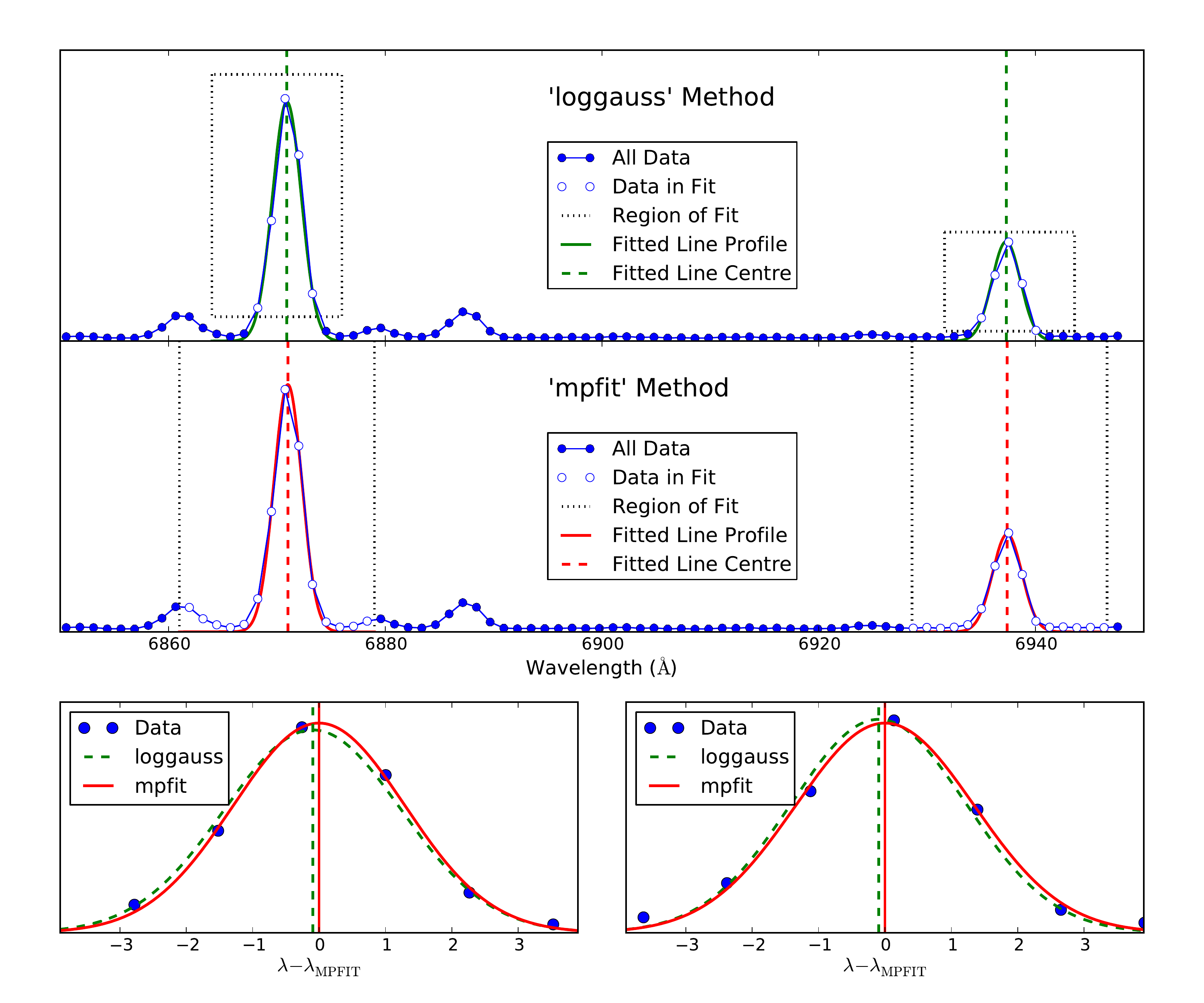}
\end{center}
\caption{Top: Fits to several arc lamp emission lines using the 'loggauss' method. Data points near a line peak with flux values exceeding 10\% of the peak flux (outlined by black dotted boxes) are fitted as a Gaussian in logarithmic flux space. Middle: Fits to the same lines using the 'mpfit' method, which utilizes {\em all} nearby data points including those with near-zero flux. Bottom Panels: Comparison of the line profiles fitted with the two methods ('loggauss' -- dashed green; 'mpfit' -- solid red), alone with their fitted centres (vertical lines). Data points are shown as blue circles.}
\label{fig:linefit}
\end{figure}

Once potential emission lines have been identified, their {\em exact} centers must be measured. \pywifes\ provides two algorithms which can be chosen by setting the 'find\_method' keyword of the \path{derive_wifes_wave_solution} routine. The first, called 'loggauss', is fast but less accurate, and is intended for rapid wavelength solution fits. It fits the logarithm of the emission line profile as a quadratic function (equivalent to a Gaussian in linear flux) but naturally must excise the pixels with zero flux from the fit. The second technique, called 'mpfit', is somewhat slower but extremely accurate, and should be considered the standard for science-grade data reduction. This method performs a Gaussian fit to the emission line flux profile using the \path{mpfit} Python package \citep[a Python port of the IDL implementation of the non-linear least squares fitting program MINPACK][]{markwardt09}. Though this procedure takes up to thirty minutes on a standard laptop, it is designed to naturally take advantage of parallel processing capabilities and generally takes about two minutes on the Linux clusters at RSAA. A visual comparison of these two methods is presented in Figure~\ref{fig:linefit}, which illustrates the sub-pixel errors in line centres found using the 'loggauss' method as compared to the more robust 'mpfit' method.

Once a series of emission lines have been identified in the data, the true wavelength of the atomic transition producing that line must be identified. Classically, this process is done interactively where an observer must visually inspect the arc lamp spectrum and compare it to a reference spectrum where lines have been identified. After identifying a few lines in the center of the longslit, a low order polynomial fit to the dispersion relation (i.e., wavelength versus pixel number) is used to automatically identify the remaining lines, and these lines are cross-identified in all remaining rows of the data. This visual inspection and line identification procedure was deemed to be cumbersome, especially given the excellent wavelength stability of the \wifes\ instrument (thermal drift of less than a few angstroms).

In practice, emission lines in \wifes\ arc lamp spectra are associated with a reference wavelength in a multi-step process. First a baseline guess for the wavelength at each found emission line position (in $x,y$ on the detector) is calculated from a known reference wavelength solution. A coherent shift of the wavelength guess for each row is applied by performing a cross-correlation of the spectrum for that row with a reference spectrum (this is optional but strongly recommended). Next, a comparison of the found line wavelengths to reference line wavelengths is performed (for each detector row), so that each found line is assigned a reference if and only if that reference wavelength is the closest wavelength in the reference list {\em and} the found wavelength is the closest found line to that reference line (i.e., a nearest neighbors requirement from both the found and reference lists). A baseline cut is applied at this stage so that associated line pairs with highly discrepant wavelength values (default is 5\AA) are rejected as false associations. 

Finally, once lines have been reliably measured and associated with a reference wavelength, the fitting of the wavelength solution proceeds. The technique employed in \pywifes\ is built on a global model of the instrument optics, which we describe in the following Section.

\subsection{Implementing the \wifes\ Optical Model}
\label{sec:optical_model}
\begin{figure}[ht!]
\begin{center}
\includegraphics[width=0.45\textwidth]{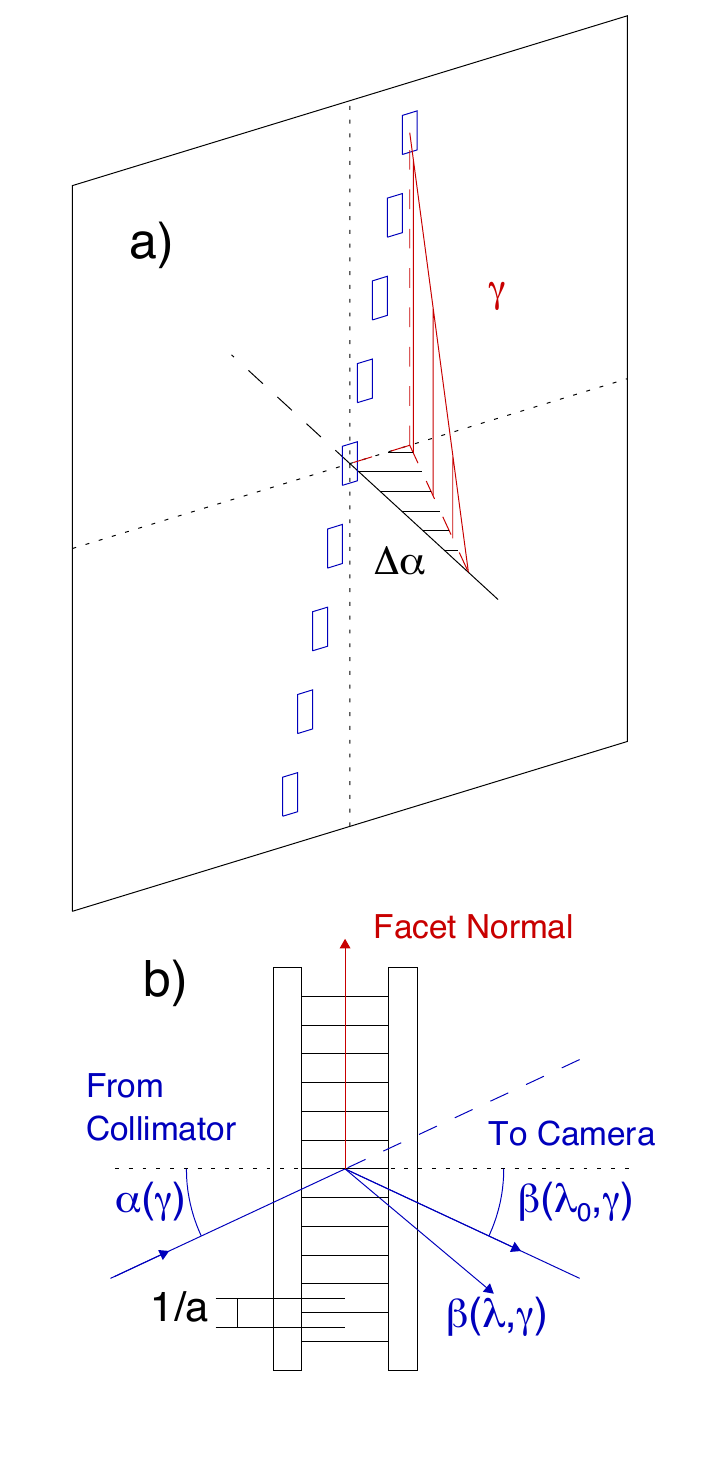}
\end{center}
\caption{Top: Schematic diagram of grating geometry, defining pertinent parameters of the grating equation. Bottom: Geometric distribution of slitlet images as passed through VPH gratings (after being split on the image slicer and reflected by other optics).}
\label{fig:grating}
\end{figure}

The manner in which light of different wavelengths is dispersed within the \wifes\ instrument can be described analytically using the geometric optics of VPH gratings \citep{barden00}. Light from all spatial positions in the \wifes\ FOV passes through the same VPH grating, but light from each spaxel has a different angle of incidence and physical point of entry due to the slit geometry from the image slicer (Fig~\ref{fig:grating}a). The slicer separates the slitlets by reflecting each spatial slice with a differential in the vertical plane – spreading the light out into a pseudo-slit. Each slitlet is also offset in the horizontal direction by widths of the neighboring slits producing the staircase slit geometry. Slitlets are further offset in the vertical direction by their own length to allow the necessary storage space on the CCD for nod-and-shuffle observations.

The light from each slitlet is dispersed following the grating equation:
\begin{equation}
  m\lambda = a \cdot \left[\sin(\alpha)+\sin(\beta)\right] \cdot \cos(\gamma)
\end{equation}
with a grating line density $a$ lines per mm. The off-axis angle, $\gamma$, introduces additional optical path difference for off-axis slits, resulting in significant spectral curvature towards the edge of the CCD. The staircase slit geometry also results in a slit-number dependence for angle of incidence $\alpha$ introducing a shift to the diffraction pattern with slit-number in addition to the classic spectral curvature.

The \wifes\ optical model is described by a total of 43 parameters. The 18 primary parameters include values such as the grating line density and blaze wavelength, the primary angle of incidence of the optical beam, the centre of the beam on the detector (in $x$ and $y$), radial distortion terms, focal length of the camera, tilt of the detector (in $x$ and $y$), as well as other geometric terms. In addition to these, each of the 25 slitlets has some very small deviation (which we label $\Delta\alpha_i$) from uniform angular separation $\alpha$ between adjacent slitlets, caused by limits in the manufacturing precision of the image slicer. Though incredibly small ($\lesssim10^{-5}$ radians), these offsets produce a coherent shift in wavelength of order 0.05\AA\ for each slitlet.

\begin{figure*}[ht!]
\begin{center}
\includegraphics[width=0.90\textwidth]{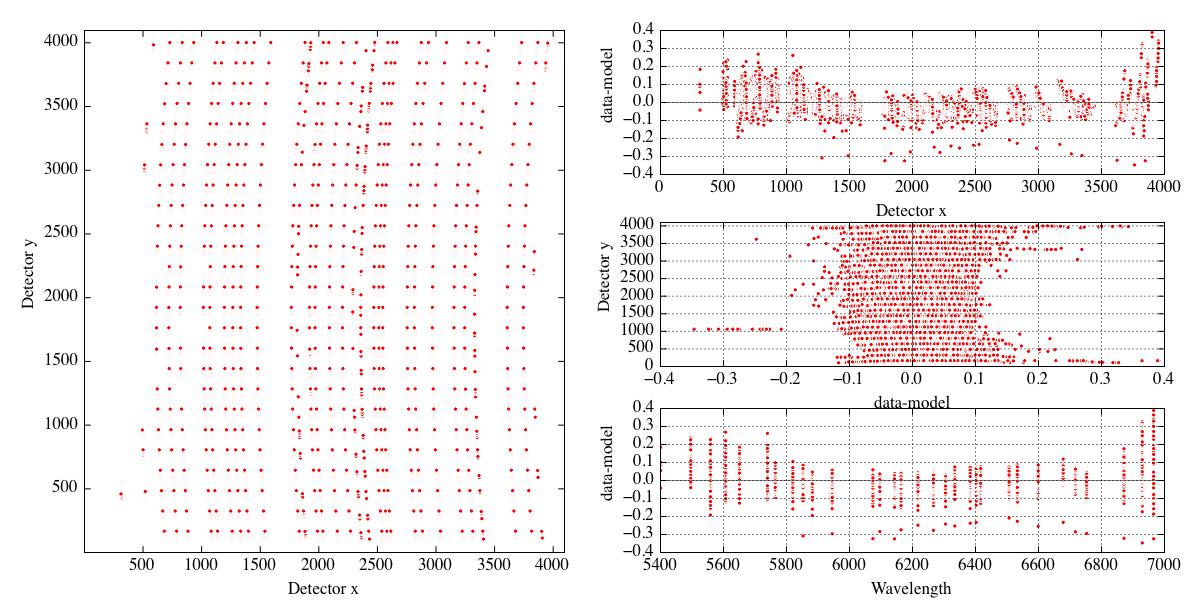}
\end{center}
\caption{Output from the \pywifes\ optical model wavelength solution for \wifes, using a Ne-Ar arc lamp exposure with the R7000 grating. Left: All 19539 found arc lines in the detector plane, illustrating the curved shape of the \wifes\ wavelength solution. Right Residuals of fitted emission line centres from prediction for the best fit optical model prediction, as a function of detector x (top), detector y (middle), and wavelength (bottom).}
\label{fig:optical_model}
\end{figure*}

The 43 \wifes\ optical model parameters are fitted by minimising the residual offset between the optical model wavelengths and reference wavelengths for the successfully found emission lines. In practice this is done in a series of steps which each fit for a specific group of optical model parameters (e.g., the 25 $\Delta\alpha_i$ values are fitted together as the final group of parameters). Residuals from this optical model method are typically of order 0.05\AA\ for the R=7000 gratings, and 0.1\AA\ for the R=3000 gratings. Figure~\ref{fig:optical_model} shows the final diagnostic plots for the optical model wavelength solution fit for a R7000 grating Ne-Ar arc lamp exposure. Structure in the residuals is largely removed, with a final residual dispersion of 0.053\AA.

In principle, many of the parameters in the \wifes\ optical model should be constant over time (the $\Delta\alpha_i$ values, the beam centre and detector tilt, detector focal length, etc.). Analysis of large volumes of \wifes\ data is currently underway to determine the stability of these parameters, and future versions of \pywifes\ will likely reflect knowledge of the static optical model parameters.

\subsection{Stability of the Wavelength Solution}
\label{sec:lam_stab}
\begin{figure*}[ht!]
\begin{center}
\includegraphics[width=0.95\textwidth]{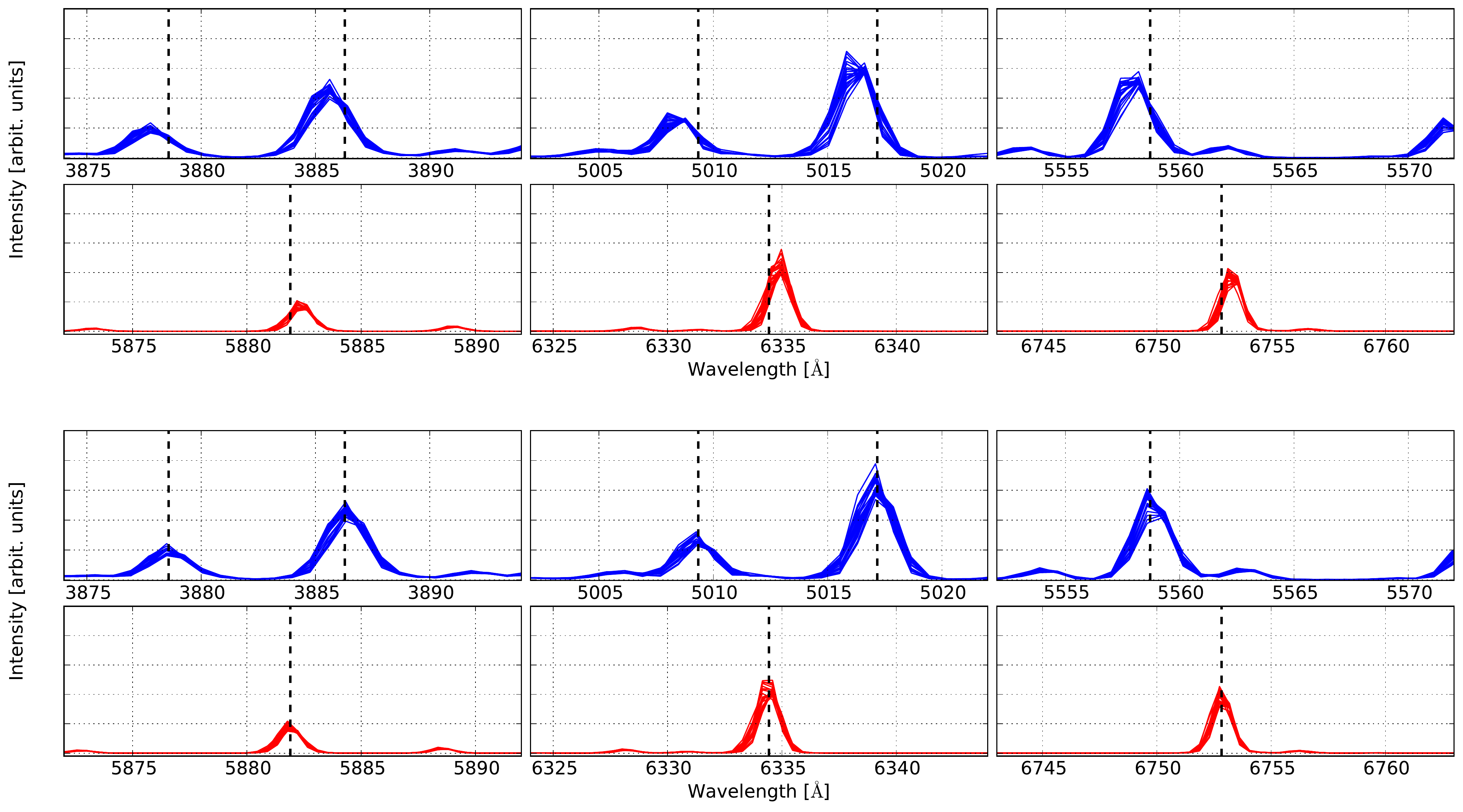}
\end{center}
\caption{Top: B3000 and R7000 Ne-Ar arc lamp spectra observed at the end of the night on 16th August 2012, and reduced using an Ne-Ar arc lamp frame acquired in the afternoon ($\sim$8 hours earlier). The vertical dashed lines denote the reference position of the arc lines. Each panel spans 20 \AA. Bottom: idem, but with the wavelength solution derived from the late-night Ne-Ar arc itself. The shifts visible in the top plots highlight the importance of taking arc frames at regular interval during an observing run to account for the temperature variations inside WiFeS.}
\label{fig:lam_drift}
\end{figure*}

\wifes\ is located on the Nasmyth focus of the 2.3m ANU telescope. This design ensures that the instrument is subject to a constant gravity vector, and makes it a very stable spectrograph overall. \wifes\ can however be subject to temperatures changes of the order of $\sim$5-15 degrees during an observing night. When subject to a change in temperature, the \wifes\ gratings will expand or contract, increasing or decreasing their resolution. To account for and correct this effect, observers are required to acquire arc frames during the night to monitor the changes of the wavelength solution. In Figure~\ref{fig:lam_drift}, we illustrate the effect of temperature drift on the wavelength solution.

In the top two panels, we show the spectra of a Ne-Ar arc lamp observed with the B3000 and R7000 gratings. We reduce the arc lamp exposure similarly to a science frame using \pywifes, and plot the 25 spectra extracted along the middle slice (along the y direction) of the final data cube. The arc frame has been acquired at the end of a winter observing night on 16th August 2012. The wavelength solution is derived from a similar arc exposure acquired early in the preceding afternoon. The reference wavelengths for the different arc lines are marked with vertical dashed lines. As expected, using an arc far-away from a given observation results in a bad wavelength calibration of the data set, with errors of the order of 1-2\AA (typically 40-60 km/s). We note that the error results in a blueshift of the spectra for the blue frame, and a redshift of the spectra for the red frame. In the bottom plot, we show the same Ne-Ar spectra, but using the frame itself to derive the wavelength solution. As expected, the arc lines are in this case in perfect agreement with their reference wavelengths. We urge \wifes\ observers to keep this point in mind when planning their observing run, and to take arc frames at regular time interval to monitor the temperature changes of the grating. Even if one's science goals do not require an absolute wavelength calibration, regular arc frames are still required to obtain a wavelength solution consistent between the red and blue frames. Our experience suggests that an interval of $\sim$1 hour between arcs is a reasonable choice, but we leave it to every observer to decide which cadence is most appropriate to their science goals.

\section{Data Reduction Scripting and Metadata}
\label{sec:top_level}
\pywifes\ was originally designed to be a generic toolkit for \wifes\ data reduction which could be scripted to match the user's desired reduction procedure. Toward this end, we designed a metadata structure and reduction script format which has become the default data reduction script included in the \pywifes\ package. The general principle underlying our preferred data reduction scheme is that the user provides the context for the data they wish to process in a generic way which can be stored in the metadata structure. That abstract structure can then be used as input to the data reduction script, which interprets the metadata structure in a completely generic fashion. This allows new data sets to be interchangeable to a degree that reduction procedures can be repeated on different data sets without requiring the user to manually specify names of data files at each step of the reduction. In this Section we first describe the default metadata structure, then briefly outline the operation of the default reduction script. A detailed walkthrough of how to generate the metadata and run the reduction script is provided on the \pywifes\ Wiki.

\subsection{Default \pywifes\ Metadata Structure}
\label{sec:metadata}
High-level information characterizing important properties of data are typically referred to as {\em metadata}. For spectroscopic observations such as those collected with \wifes, examples of metadata include the observation type, the target being observed, or whether multiple exposures should be associated with one another. This information is then used to decide which data reduction procedures should be used, and is perhaps most commonly handled manually. Because spectroscopic data reduction often follows a repeated pattern, we desired an abstract metadata structure which could be assigned to a particular set of data then handled generically by a standard data reduction script.

In practice, the \pywifes\ metadata structure for a given set of data is stored in a Python dictionary which is saved in a Python pickle file. This dictionary is defined in the scripts \path{save_blue_metadata.py} and \path{save_red_metadata.py} (the metadata structures must be defined separately for the red and blue cameras), which we will call the ``metadata saving'' scripts. These are included in the \pywifes\ distribution and each script simply defines the metadata dictionary and saves it to a pickle file.

The metadata dictionary has several keywords which store information about the various types of calibration images, as well as science observation groups, and observations of standard stars. Generic instrumental calibration files -- such as flat field observations, arc lamp exposures, and bias frames -- which can be applied to calibrating all science and standard star data are stored in simple lists. These files are considered to be the ``master'' calibration files for a given data reduction run, and are applied in the calibration of all files unless specific calibration files are specified for a given observation.

Science and standard star observations are each stored as a list of Python dictionaries. Each dictionary describes an observation group, including all science frames which are ultimately co-added, as well as associated calibration frames (``local'' calibrations) taken explicitly for use in calibrating that particular science target (e.g., a bias frame and arc frame taken immediately after observing a science target). Each keyword of the dictionary is a list, with the 'sci' keyword containing a list of all observations to be co-added as the final science frame, the 'sky' keyword containing a list (typically of length 1) of sky frames to be subtracted during data reduction, 'arc' containing the ``local'' arc frame, and so on. Standard star dictionaries also have the special keyword 'type' whose value is a list consisting of one or both of the strings 'flux' and 'telluric', used to distinguished if that star is used in the flux calibration and/or telluric correction procedures.

\begin{figure}[h!]
\begin{center}
\includegraphics[width=0.45\textwidth]{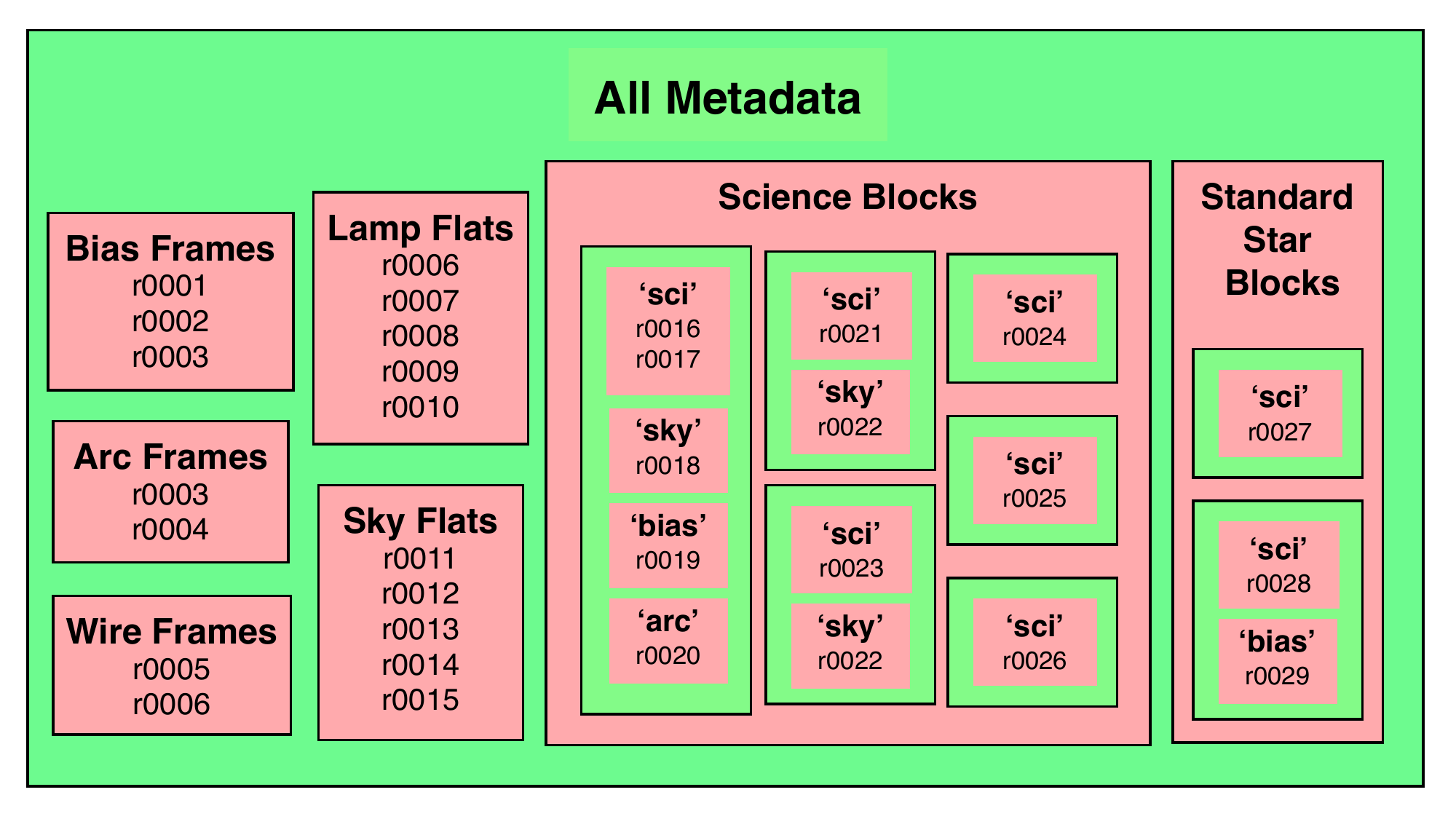}
\end{center}
\caption{Schematic diagram of the default \pywifes\ metadata structure. Python dictionaries are green, and Python lists are red. Names or descriptions of the lists or dictionaries are in bold face, while place face text denotes example file names. Master calibration files are stored in Python lists within the metadata dictionary, which the science and standard star observations are lists of dictionaries. Science observation block dictionaries must at minimum have the 'sci' keyword defined, but may also have calibration and sky frames associated to them as appropriately named lists in the observation block dictionaries.}
\label{fig:metadata}
\end{figure}

For example, if the only science observation is a pair of science frames followed by a sky frame, then a bias and an arc frame, the pertinent part of red metadata saving script would appear as this:
\begin{verbatim}
   sci_obs = [
       {'sci'  : ['r0001', 'r0002'],
        'sky'  : ['r0003'],
        'bias' : ['r0004'],
        'arc'  : ['r0005'],
        },
   ]
\end{verbatim}
This illustrates the naming convention employed in the metadata structure, where the observation root name is used to identify the data so that \path{'r0001'} refers to the observation whose raw data is in the file ``\path{r0001.fits}''. At the end of the metadata script, the list \path{'sci_obs'} is saved as the 'sci' keyword of the metadata dictionary. A summary schematic diagram of the \pywifes\ metadata structure is shown in Figure~\ref{fig:metadata}.

The metadata structure is saved to a Python pickle file by running the \path{save_red_metadata.py} script from the command line, and this pickle file is passed as input to the data reduction script. The metadata saving script is a critical part of the data reduction scheme, and should be carefully prepared by the user before each reduction.

For convenience, we include with \pywifes\ a Python script which inspects all FITS files in a directory and attempts to automatically generate the metadata saving scripts (for blue and red). We caution that this script is very simplistic, and identifies calibrations by the \path{'IMAGETYP'} header keyword set by the \wifes\ controller. All calibration files are by default assigned to the ``master'' calibration lists, and any calibrations which the user took for specific science frames should be assigned by the user to those frames as in the example provided above. The \pywifes\ metadata structure is designed to be as abstract as possible to enable ease of data reduction, but requires careful construction of the metadata by the user in order to ensure data reduction is performed correctly.

\subsection{The \pywifes\ Data Reduction Script}
\label{sec:script}
The \pywifes\ data reduction scripts \path{reduce_blue_data.py} and \path{reduce_red_data.py} take a \pywifes\ metadata file as input and perform the data reduction procedures outlined by the user. The ``standard'' version of these scripts are included in the \pywifes\ distribution, but can be edited to add or modify data reduction steps as desired. 
In this Section we describe the default structure of the reduction script.


The data reduction procedure performed by the reduction script is defined in a Python list \path{proc_steps}, which is a list of the data reduction steps. Each step is defined by a Python dictionary which has four keywords:
\begin{itemize}
  \item \path{step} -- This is the main label for each reduction step and identifies which function (described below) should be called. For example, the first step ``overscan\_sub'' calls the function \path{run_overscan_sub}, which in turn calls the appropriate \pywifes\ functions to operate on the data.
  \item \path{run} -- A simple Python Boolean (\path{True} or \path{False}) which sets whether that step in the reduction is run. This keyword is useful when individual steps of the reduction need to be re-run, e.g., if the wavelength solution does not converge to the user's satisfaction. It is important to note that a step which has been set to \path{'run: False'} will be interpreted by the reduction script as having been successfully executed, and the following step will look for the output of the ``switched-off'' step. If users wish to completely remove a step from the reduction chain, it should either be commented out or deleted.
  \item \path{suffix} -- This defines how the output of the reduction step will be labeled. For example, an image whose root name is 'r0001' (as defined in the metadata file) and is processed through the first reduction step with \path{'suffix':'01'} would have output ``r0001.p01.fits'' where the 'p' indicates the file has been through the processing step labeled by '01'. The reduction script knows to look for the suffix of the previous step to define the input image to that step. Following the same example, 'r0001' passed to the next step with \path{'suffix':'02'} would result in that step looking for ``r0001.p01.fits'', processing it, and saving the output in ``r0001.p02.fits''.
  \item \path{args} -- This keyword is set as a dictionary of arguments which should be passed to the main function being called in that data reduction step. For example, in the ``cube\_gen'' step which generates final data cubes, one can set the minimum wavelength of the cubes by setting the \path{args} keyword as \path{'args':\{'wmin_set':5400.0\}}. This will pass the argument on so that every time \path{generate_wifes_cube} is called in this step, it will be called with the keyword argument \path{wmin_set = 5400.0}. This functionality allows the user to adjust the output of the \pywifes\ functions easily from the processing step list.
\end{itemize}

Each reduction step calls a ``master reduction function'' (such as \path{run_overscan_sub} in the example above) which is defined in the reduction script (which can be modified by the user). These functions {\em must} be defined to take as input the metadata structure (the Python dictionary described in Section~\ref{sec:metadata}), the suffix from the previous reduction step (\path{prev_suffix}), the suffix to be used in output from the current step (\path{curr_suffix}), and finally the additional arguments to be passed the pertinent \pywifes\ functions. A loop at the end of the reduction script automatically parses the processing step list (\path{proc_steps}) and calls the master reduction functions with the correct suffixes.

Each master reduction function parses the metadata in a specific way and calls the pertinent functions from \pywifes\ to perform that reduction step. For example, the \path{run_overscan_sub} function calls \path{pywifes.subtract_overscan} on all raw frames, \path{run_obs_coadd} calls \path{pywifes.imarith_mef} on all groups of science and standard star frames, and \path{run_extract_stars} calls \path{pywifes.extract_wifes_stdstar} on all standard star data cubes. 

Some steps in the processing list do not perform operations on all data, but instead prepare master calibration files such as the flatfield response function or the wavelength solution. The master reduction function which generates these master calibration files also checks to see if any science observation groups require the use of ``local'' calibration files, and derives the appropriate calibrations from those files as well. The names of the master files are defined early in the reduction script (and again can be changed as desired by the user) and are global variables which are referenced when calibrations must be applied in the processing steps, but use of local calibrations is also handled automatically. For example, in the data cube generation step, science frames are transformed into data cubes using the master wire and master wavelength solution by default, but using the local wavelength solution for observation groups where a ``local arc'' was defined.

The full list of the default processing steps is:
\begin{description}
  \item[Step 00] Overscan subtraction (must happen before {\em any} other step).
  \item[Step 01] Bad pixel repair.
  \item[({\em CAL})] Creation of ``superbias'', and bias model from superbias and any local biases.
  \item[Step 02] Bias frame subtraction.
  \item[({\em CAL})] Creation of ``superflats'', both lamp and sky.
  \item[({\em CAL})] Measurement of the slitlet profiles.
  \item[({\em CAL})] Desag (optional)
  \item[({\em CAL})] Flat frame cleanup.
  \item[({\em CAL})] MEF flat creation, both lamp and sky.
  \item[Step 03] MEF creation for all frames.
  \item[({\em CAL})] Derivation of the wavelength solution.
  \item[({\em CAL})] Derivation of the wire (spatial) solution.
  \item[({\em CAL})] Derivation of the flatfield response function (requires wavelength solution).
  \item[Step 04] Cosmic ray rejection.
  \item[Step 05] Subtraction of sky frames.
  \item[Step 06] Co-adding of frames in observation groups.
  \item[Step 07] Application of flatfield response function.
  \item[Step 08] Generation of data cubes.
  \item[({\em CAL})] Extraction of uncalibrated standard star spectra.
  \item[({\em CAL})] Derivation of flux calibration solution from standard star spectra.
  \item[Step 09] Flux calibration of all data cubes.
  \item[({\em CAL})] Extraction of calibrated, but not telluric-corrected, standard star spectra.
  \item[({\em CAL})] Derivation of telluric correction from standard star spectra.
  \item[Step 10] Correction of telluric absorption.
  \item[Step 11] Reformatting into 3D data cube format.
\end{description}
Steps which perform processing operations on all science frames (and other frames requiring that reduction step) are labeled by number in the above list, while steps labeled as ({\em CAL}) denote processing steps which prepare or operate on master calibration files.

We also note that nod-and-shuffle (``N+S'') frames are handled automatically in the \pywifes\ reduction scripts. Special treatment of N+S data needs to occur when the MEF files are first created, when subtraction of sky frames occurs, and any intermediate steps between. In the default scheme described above, MEF files are created at the 'p03' step, and N+S frames are automatically processed with the N+S version of the slitlet separation function. For example, if the raw \wifes\ image was ``r0001.fits'', the N+S version of this step would save the object pixels as the MEF file ``r0001.p03.fits'' and the nodded sky pixels are saved as the MEF file ``r0001.s03.fits''. By default, these files are both processed in the cosmic ray step to produce ``r0001.p04.fits'' and ``r0001.s04.fits'' files, then in the sky subtraction step the nodded sky pixels in ``r0001.s04.fits'' are subtracted from the object pixels in ``r0001.p04.fits'' to produce the sky-subtracted ``r0001.p05.fits''.  All of these steps are executed automatically without the user needing to identify N+S frame, as these are identified from the 'WIFESOBS' FITS header keyword.

Finally, we note that all \pywifes\ data processing functions are designed to work seamlessly with \wifes\ data taken with detector binning, or in half-frame readout mode. However, the data reduction script and metadata creation script are not currently designed to handle multiple observing modes at the same time (though that functionality is planned for future upgrades). For example, attempting to subtract a full-frame bias from a half-frame image will cause the reduction to fail. We caution that these failures modes are not Python exceptions built specifically into \pywifes, but rather are the results of normal Python errors being raised (e.g., due to incompatibility of data size). This means that it is possible to perform reduction steps which are wrong but do not raise Python errors, such as subtracting a full-frame bias with binning of 2 from a half-frame image with binning of 1, or flux calibrating R7000 data with a calibration solution derived for the R3000 grating. Again, users should be careful in compiling their metadata so that only compatible data are being sent to the data reduction script.

\section{Conclusion}
\label{sec:conclusion}
We presented \pywifes, a new rapid data reduction pipeline for the \wifes\ instrument. \pywifes\ consists of a series of data processing functions written in Python which utilise existing functionality from the standard Python modules NumPy, SciPy, and PyFITS. \pywifes\ does not make any function calls outside of Python, and thus requires no special system-dependent software installation nor does it require other astronomical data processing environments such as IRAF.

Data processing functions in \pywifes\ are designed to operate on FITS format data as both input and output. This means data processing naturally occurs as a modular process, and the order of data reduction steps can be easily adjusted by constructing Python scripts. More importantly, the scriptable nature of \pywifes\ enables repeat processing to accommodate upgrades to the reduction code or to process new data in the same way as previous data sets. \pywifes\ data reduction functions were also carefully constructed to define instrument characteristics with abstract variables wherever possible so that the code could potentially be adapted for use in reducing data from other image-slicing integral field spectrographs.

Beyond the functional data reduction procedures defined in the core \pywifes\ modules, the reduction pipeline also includes a default metadata format and reduction script favored by the developers. The main goal of this structure is to decouple the assignment of calibration or processing roles to specific files from the data reduction procedures performed once those roles are assigned. Specifically, if an observer uses arc, flat, and bias frames in the same data reduction roles for every night of observing, the observer should not have to repeat or edit calls to data reduction procedures simply to reflect the different file names from different nights. Our default \pywifes\ metadata structure accomplishes this goal by storing the observation metadata in an abstractly labeled Python dictionary. For example, all bias frames for a night are stored in a Python list within the metadata dictionary, and the data reduction script operates on bias frames when needed by iterating through members of that list (without requiring knowledge of the exact file name). Metadata for science observations are stored as a list of Python dictionaries containing lists of the science frames themselves and any associated calibration frames, so that calibration data needing to be applied to a specific frame can be easily tracked.

While this paper outlines the official release version of \pywifes, future improvements to the pipeline will be implemented as needed and released from the RSAA \pywifes\ Wiki. The current version of the pipeline already accounts for all standard integral field data reduction procedures, as well as several sophisticated algorithms designed to account for realistic behavior of the \wifes\ instrument. These include the scattered light flatfield correction procedures, which result in excellent spatial flatfielding, and the global wavelength solution based on the instrument optics. Future data reduction improvements based on better understanding of the instrument behavior will be incorporated when identified, and improvements in data reduction scripting will be released with future developments as well.

%
\acknowledgments
We thank the many \wifes\ users who used \pywifes\ in its early ``beta testing'' stages and contributed feedback and data which significantly improved the quality of the pipeline. These contributors include: Mike Bessell, Raul Cacho, Rebecca Davies, Catherine de Burgh-Day, I-Ting Ho, Wolfgang Kerzendorf, Sarah Leslie, Melissa Ness, David Nicholls, Chris Owen, Sinem Ozbilgen, Matthew Satterthwaite, Marja Seidel, Tiantian Yuan, and Jabran Zahid.  We also thank the referee Rog\'erio Riffel for careful reading of the text and thoughtful comments.
This research was conducted by the Australian Research Council Centre of Excellence for All-sky Astrophysics (CAASTRO), through project number CE110001020.


%
 \bibliographystyle{spr-mp-nameyear-cnd}  
 \bibliography{pywifes_paper}             

\begin{thebibliography}{24}
\ifx \bisbn   \undefined \def \bisbn  #1{ISBN #1}\fi
\ifx \binits  \undefined \def \binits#1{#1} \fi
\ifx \bauthor  \undefined \def \bauthor#1{#1} \fi
\ifx \batitle  \undefined \def \batitle#1{#1} \fi
\ifx \bjtitle  \undefined \def \bjtitle#1{#1}\fi
\ifx \bvolume  \undefined \def \bvolume#1{\textbf{#1}}\fi
\ifx \byear  \undefined \def \byear#1{#1} \fi
\ifx \bissue  \undefined \def \bissue#1{#1} \fi
\ifx \bfpage  \undefined \def \bfpage#1{#1} \fi
\ifx \blpage  \undefined \def \blpage #1{#1} \fi
\ifx \burl  \undefined \def \burl#1{\textsf{#1}} \fi
\ifx \doiurl  \undefined \def \doiurl#1{\textsf{#1}} \fi
\ifx \betal  \undefined \def \betal{\textit{et al.}} \fi
\ifx \binstitute  \undefined \def \binstitute#1{#1} \fi
\ifx \binstitutionaled  \undefined \def \binstitutionaled#1{#1} \fi
\ifx \bctitle  \undefined \def \bctitle#1{#1} \fi
\ifx \beditor  \undefined \def \beditor#1{#1} \fi
\ifx \bpublisher  \undefined \def \bpublisher#1{#1} \fi
\ifx \bbtitle  \undefined \def \bbtitle#1{#1} \fi
\ifx \bedition  \undefined \def \bedition#1{#1} \fi
\ifx \bseriesno  \undefined \def \bseriesno#1{#1} \fi
\ifx \blocation  \undefined \def \blocation#1{#1} \fi
\ifx \bsertitle  \undefined \def \bsertitle#1{#1} \fi
\ifx \bsnm \undefined \def \bsnm#1{#1} \fi
\ifx \bsuffix \undefined \def \bsuffix#1{#1} \fi
\ifx \bparticle \undefined \def \bparticle#1{#1} \fi
\ifx \barticle \undefined \def \barticle#1{#1} \fi
\ifx \bconfdate \undefined \def \bconfdate #1{#1} \fi
\ifx \botherref \undefined \def \botherref #1{#1} \fi
\ifx \url \undefined \def \url#1{\textsf{#1}} \fi
\ifx \bchapter \undefined \def \bchapter#1{#1} \fi
\ifx \bbook \undefined \def \bbook#1{#1} \fi
\ifx \bcomment \undefined \def \bcomment#1{#1} \fi
\ifx \oauthor \undefined \def \oauthor#1{#1} \fi
\ifx \citeauthoryear \undefined \def \citeauthoryear#1{#1} \fi
\ifx \endbibitem  \undefined \def \endbibitem {}\fi
\ifx \bconflocation  \undefined \def \bconflocation#1{#1} \fi
\ifx \arxivurl  \undefined \def \arxivurl#1{\textsf{#1}} \fi

\bibitem[\protect\citeauthoryear{{Bakos} et~al.}{2013}]{bakos13}
\begin{barticle}
\bauthor{\bsnm{{Bakos}}, \binits{G.{\'A}.}},
\bauthor{\bsnm{{Csubry}}, \binits{Z.}},
\bauthor{\bsnm{{Penev}}, \binits{K.}},
\bauthor{\bsnm{{Bayliss}}, \binits{D.}},
\bauthor{\bsnm{{Jord{\'a}n}}, \binits{A.}},
\bauthor{\bsnm{{Afonso}}, \binits{C.}},
\bauthor{\bsnm{{Hartman}}, \binits{J.D.}},
\bauthor{\bsnm{{Henning}}, \binits{T.}},
\bauthor{\bsnm{{Kov{\'a}cs}}, \binits{G.}},
\bauthor{\bsnm{{Noyes}}, \binits{R.W.}},
\bauthor{\bsnm{{B{\'e}ky}}, \binits{B.}},
\bauthor{\bsnm{{Suc}}, \binits{V.}},
\bauthor{\bsnm{{Cs{\'a}k}}, \binits{B.}},
\bauthor{\bsnm{{Rabus}}, \binits{M.}},
\bauthor{\bsnm{{L{\'a}z{\'a}r}}, \binits{J.}},
\bauthor{\bsnm{{Papp}}, \binits{I.}},
\bauthor{\bsnm{{S{\'a}ri}}, \binits{P.}},
\bauthor{\bsnm{{Conroy}}, \binits{P.}},
\bauthor{\bsnm{{Zhou}}, \binits{G.}},
\bauthor{\bsnm{{Sackett}}, \binits{P.D.}},
\bauthor{\bsnm{{Schmidt}}, \binits{B.}},
\bauthor{\bsnm{{Mancini}}, \binits{L.}},
\bauthor{\bsnm{{Sasselov}}, \binits{D.D.}},
\bauthor{\bsnm{{Ueltzhoeffer}}, \binits{K.}}:
\bjtitle{\pasp}
\bvolume{125},
\bfpage{154}
(\byear{2013}).
\arxivurl{1206.1391}.
doi:\doiurl{10.1086/669529}
\end{barticle}
\endbibitem

\bibitem[\protect\citeauthoryear{{Barden} et~al.}{2000}]{barden00}
\begin{barticle}
\bauthor{\bsnm{{Barden}}, \binits{S.C.}},
\bauthor{\bsnm{{Arns}}, \binits{J.A.}},
\bauthor{\bsnm{{Colburn}}, \binits{W.S.}},
\bauthor{\bsnm{{Williams}}, \binits{J.B.}}:
\bjtitle{\pasp}
\bvolume{112},
\bfpage{809}
(\byear{2000}).
doi:\doiurl{10.1086/316576}
\end{barticle}
\endbibitem

\bibitem[\protect\citeauthoryear{{Bayliss} et~al.}{2013}]{bayliss13}
\begin{botherref}
\oauthor{\bsnm{{Bayliss}}, \binits{D.}},
\oauthor{\bsnm{{Zhou}}, \binits{G.}},
\oauthor{\bsnm{{Penev}}, \binits{K.}},
\oauthor{\bsnm{{Bakos}}, \binits{G.}},
\oauthor{\bsnm{{Hartman}}, \binits{J.}},
\oauthor{\bsnm{{Jord{\'a}n}}, \binits{A.}},
\oauthor{\bsnm{{Mancini}}, \binits{L.}},
\oauthor{\bsnm{{Mohler}}, \binits{M.}},
\oauthor{\bsnm{{Suc}}, \binits{V.}},
\oauthor{\bsnm{{Rabus}}, \binits{M.}},
\oauthor{\bsnm{{B{\'e}ky}}, \binits{B.}},
\oauthor{\bsnm{{Csubry}}, \binits{Z.}},
\oauthor{\bsnm{{Buchhave}}, \binits{L.}},
\oauthor{\bsnm{{Henning}}, \binits{T.}},
\oauthor{\bsnm{{Nikolov}}, \binits{N.}},
\oauthor{\bsnm{{Cs{\'a}k}}, \binits{B.}},
\oauthor{\bsnm{{Brahm}}, \binits{R.}},
\oauthor{\bsnm{{Espinoza}}, \binits{N.}},
\oauthor{\bsnm{{Noyes}}, \binits{R.}},
\oauthor{\bsnm{{Schmidt}}, \binits{B.}},
\oauthor{\bsnm{{Conroy}}, \binits{P.}},
\oauthor{\bsnm{{Wright}}, \binits{D.}},
\oauthor{\bsnm{{Tinney}}, \binits{C.}},
\oauthor{\bsnm{{Addison}}, \binits{B.}},
\oauthor{\bsnm{{Sackett}}, \binits{P.}},
\oauthor{\bsnm{{Sasselov}}, \binits{D.}},
\oauthor{\bsnm{{L{\'a}z{\'a}r}}, \binits{J.}},
\oauthor{\bsnm{{Papp}}, \binits{I.}},
\oauthor{\bsnm{{S{\'a}ri}}, \binits{P.}}:
ArXiv e-prints
(2013).
\arxivurl{1306.0624}
\end{botherref}
\endbibitem

\bibitem[\protect\citeauthoryear{{Bessell}}{1999}]{bessell99}
\begin{barticle}
\bauthor{\bsnm{{Bessell}}, \binits{M.S.}}:
\bjtitle{\pasp}
\bvolume{111},
\bfpage{1426}
(\byear{1999}).
doi:\doiurl{10.1086/316454}
\end{barticle}
\endbibitem

\bibitem[\protect\citeauthoryear{{Bessell}}{2005}]{bessell05}
\begin{barticle}
\bauthor{\bsnm{{Bessell}}, \binits{M.S.}}:
\bjtitle{\araa}
\bvolume{43},
\bfpage{293}
(\byear{2005}).
doi:\doiurl{10.1146/annurev.astro.41.082801.100251}
\end{barticle}
\endbibitem

\bibitem[\protect\citeauthoryear{{Buton} et~al.}{2013}]{buton13}
\begin{barticle}
\bauthor{\bsnm{{Buton}}, \binits{C.}},
\bauthor{\bsnm{{Copin}}, \binits{Y.}},
\bauthor{\bsnm{{Aldering}}, \binits{G.}},
\bauthor{\bsnm{{Antilogus}}, \binits{P.}},
\bauthor{\bsnm{{Aragon}}, \binits{C.}},
\bauthor{\bsnm{{Bailey}}, \binits{S.}},
\bauthor{\bsnm{{Baltay}}, \binits{C.}},
\bauthor{\bsnm{{Bongard}}, \binits{S.}},
\bauthor{\bsnm{{Canto}}, \binits{A.}},
\bauthor{\bsnm{{Cellier-Holzem}}, \binits{F.}},
\bauthor{\bsnm{{Childress}}, \binits{M.}},
\bauthor{\bsnm{{Chotard}}, \binits{N.}},
\bauthor{\bsnm{{Fakhouri}}, \binits{H.K.}},
\bauthor{\bsnm{{Gangler}}, \binits{E.}},
\bauthor{\bsnm{{Guy}}, \binits{J.}},
\bauthor{\bsnm{{Hsiao}}, \binits{E.Y.}},
\bauthor{\bsnm{{Kerschhaggl}}, \binits{M.}},
\bauthor{\bsnm{{Kowalski}}, \binits{M.}},
\bauthor{\bsnm{{Loken}}, \binits{S.}},
\bauthor{\bsnm{{Nugent}}, \binits{P.}},
\bauthor{\bsnm{{Paech}}, \binits{K.}},
\bauthor{\bsnm{{Pain}}, \binits{R.}},
\bauthor{\bsnm{{P{\'e}contal}}, \binits{E.}},
\bauthor{\bsnm{{Pereira}}, \binits{R.}},
\bauthor{\bsnm{{Perlmutter}}, \binits{S.}},
\bauthor{\bsnm{{Rabinowitz}}, \binits{D.}},
\bauthor{\bsnm{{Rigault}}, \binits{M.}},
\bauthor{\bsnm{{Runge}}, \binits{K.}},
\bauthor{\bsnm{{Scalzo}}, \binits{R.}},
\bauthor{\bsnm{{Smadja}}, \binits{G.}},
\bauthor{\bsnm{{Tao}}, \binits{C.}},
\bauthor{\bsnm{{Thomas}}, \binits{R.C.}},
\bauthor{\bsnm{{Weaver}}, \binits{B.A.}},
\bauthor{\bsnm{{Wu}}, \binits{C.}},
\bauthor{\bsnm{{Nearby SuperNova Factory}}}:
\bjtitle{\aap}
\bvolume{549},
\bfpage{8}
(\byear{2013}).
\arxivurl{1210.2619}.
doi:\doiurl{10.1051/0004-6361/201219834}
\end{barticle}
\endbibitem

\bibitem[\protect\citeauthoryear{{Childress} et~al.}{2013}]{childress12fr}
\begin{barticle}
\bauthor{\bsnm{{Childress}}, \binits{M.J.}},
\bauthor{\bsnm{{Scalzo}}, \binits{R.A.}},
\bauthor{\bsnm{{Sim}}, \binits{S.A.}},
\bauthor{\bsnm{{Tucker}}, \binits{B.E.}},
\bauthor{\bsnm{{Yuan}}, \binits{F.}},
\bauthor{\bsnm{{Schmidt}}, \binits{B.P.}},
\bauthor{\bsnm{{Cenko}}, \binits{S.B.}},
\bauthor{\bsnm{{Silverman}}, \binits{J.M.}},
\bauthor{\bsnm{{Contreras}}, \binits{C.}},
\bauthor{\bsnm{{Hsiao}}, \binits{E.Y.}},
\bauthor{\bsnm{{Phillips}}, \binits{M.}},
\bauthor{\bsnm{{Morrell}}, \binits{N.}},
\bauthor{\bsnm{{Jha}}, \binits{S.W.}},
\bauthor{\bsnm{{McCully}}, \binits{C.}},
\bauthor{\bsnm{{Filippenko}}, \binits{A.V.}},
\bauthor{\bsnm{{Anderson}}, \binits{J.P.}},
\bauthor{\bsnm{{Benetti}}, \binits{S.}},
\bauthor{\bsnm{{Bufano}}, \binits{F.}},
\bauthor{\bsnm{{de Jaeger}}, \binits{T.}},
\bauthor{\bsnm{{Forster}}, \binits{F.}},
\bauthor{\bsnm{{Gal-Yam}}, \binits{A.}},
\bauthor{\bsnm{{Le Guillou}}, \binits{L.}},
\bauthor{\bsnm{{Maguire}}, \binits{K.}},
\bauthor{\bsnm{{Maund}}, \binits{J.}},
\bauthor{\bsnm{{Mazzali}}, \binits{P.A.}},
\bauthor{\bsnm{{Pignata}}, \binits{G.}},
\bauthor{\bsnm{{Smartt}}, \binits{S.}},
\bauthor{\bsnm{{Spyromilio}}, \binits{J.}},
\bauthor{\bsnm{{Sullivan}}, \binits{M.}},
\bauthor{\bsnm{{Taddia}}, \binits{F.}},
\bauthor{\bsnm{{Valenti}}, \binits{S.}},
\bauthor{\bsnm{{Bayliss}}, \binits{D.D.R.}},
\bauthor{\bsnm{{Bessell}}, \binits{M.}},
\bauthor{\bsnm{{Blanc}}, \binits{G.A.}},
\bauthor{\bsnm{{Carson}}, \binits{D.J.}},
\bauthor{\bsnm{{Clubb}}, \binits{K.I.}},
\bauthor{\bsnm{{de Burgh-Day}}, \binits{C.}},
\bauthor{\bsnm{{Desjardins}}, \binits{T.D.}},
\bauthor{\bsnm{{Fang}}, \binits{J.J.}},
\bauthor{\bsnm{{Fox}}, \binits{O.D.}},
\bauthor{\bsnm{{Gates}}, \binits{E.L.}},
\bauthor{\bsnm{{Ho}}, \binits{I.-T.}},
\bauthor{\bsnm{{Keller}}, \binits{S.}},
\bauthor{\bsnm{{Kelly}}, \binits{P.L.}},
\bauthor{\bsnm{{Lidman}}, \binits{C.}},
\bauthor{\bsnm{{Loaring}}, \binits{N.S.}},
\bauthor{\bsnm{{Mould}}, \binits{J.R.}},
\bauthor{\bsnm{{Owers}}, \binits{M.}},
\bauthor{\bsnm{{Ozbilgen}}, \binits{S.}},
\bauthor{\bsnm{{Pei}}, \binits{L.}},
\bauthor{\bsnm{{Pickering}}, \binits{T.}},
\bauthor{\bsnm{{Pracy}}, \binits{M.B.}},
\bauthor{\bsnm{{Rich}}, \binits{J.A.}},
\bauthor{\bsnm{{Schaefer}}, \binits{B.E.}},
\bauthor{\bsnm{{Scott}}, \binits{N.}},
\bauthor{\bsnm{{Stritzinger}}, \binits{M.}},
\bauthor{\bsnm{{Vogt}}, \binits{F.P.A.}},
\bauthor{\bsnm{{Zhou}}, \binits{G.}}:
\bjtitle{\apj}
\bvolume{770},
\bfpage{29}
(\byear{2013}).
\arxivurl{1302.2926}.
doi:\doiurl{10.1088/0004-637X/770/1/29}
\end{barticle}
\endbibitem

\bibitem[\protect\citeauthoryear{{Dopita} et~al.}{2007}]{dopita07}
\begin{barticle}
\bauthor{\bsnm{{Dopita}}, \binits{M.}},
\bauthor{\bsnm{{Hart}}, \binits{J.}},
\bauthor{\bsnm{{McGregor}}, \binits{P.}},
\bauthor{\bsnm{{Oates}}, \binits{P.}},
\bauthor{\bsnm{{Bloxham}}, \binits{G.}},
\bauthor{\bsnm{{Jones}}, \binits{D.}}:
\bjtitle{\apss}
\bvolume{310},
\bfpage{255}
(\byear{2007}).
\arxivurl{0705.0287}.
doi:\doiurl{10.1007/s10509-007-9510-z}
\end{barticle}
\endbibitem

\bibitem[\protect\citeauthoryear{{Dopita} et~al.}{2010}]{dopita10}
\begin{barticle}
\bauthor{\bsnm{{Dopita}}, \binits{M.}},
\bauthor{\bsnm{{Rhee}}, \binits{J.}},
\bauthor{\bsnm{{Farage}}, \binits{C.}},
\bauthor{\bsnm{{McGregor}}, \binits{P.}},
\bauthor{\bsnm{{Bloxham}}, \binits{G.}},
\bauthor{\bsnm{{Green}}, \binits{A.}},
\bauthor{\bsnm{{Roberts}}, \binits{B.}},
\bauthor{\bsnm{{Neilson}}, \binits{J.}},
\bauthor{\bsnm{{Wilson}}, \binits{G.}},
\bauthor{\bsnm{{Young}}, \binits{P.}},
\bauthor{\bsnm{{Firth}}, \binits{P.}},
\bauthor{\bsnm{{Busarello}}, \binits{G.}},
\bauthor{\bsnm{{Merluzzi}}, \binits{P.}}:
\bjtitle{\apss}
\bvolume{327},
\bfpage{245}
(\byear{2010}).
\arxivurl{1002.4472}.
doi:\doiurl{10.1007/s10509-010-0335-9}
\end{barticle}
\endbibitem

\bibitem[\protect\citeauthoryear{{Filippenko}}{1982}]{filippenko82}
\begin{barticle}
\bauthor{\bsnm{{Filippenko}}, \binits{A.V.}}:
\bjtitle{\pasp}
\bvolume{94},
\bfpage{715}
(\byear{1982}).
doi:\doiurl{10.1086/131052}
\end{barticle}
\endbibitem

\bibitem[\protect\citeauthoryear{{Fraser} et~al.}{2013}]{fraser09ip}
\begin{barticle}
\bauthor{\bsnm{{Fraser}}, \binits{M.}},
\bauthor{\bsnm{{Inserra}}, \binits{C.}},
\bauthor{\bsnm{{Jerkstrand}}, \binits{A.}},
\bauthor{\bsnm{{Kotak}}, \binits{R.}},
\bauthor{\bsnm{{Pignata}}, \binits{G.}},
\bauthor{\bsnm{{Benetti}}, \binits{S.}},
\bauthor{\bsnm{{Botticella}}, \binits{M.-T.}},
\bauthor{\bsnm{{Bufano}}, \binits{F.}},
\bauthor{\bsnm{{Childress}}, \binits{M.}},
\bauthor{\bsnm{{Mattila}}, \binits{S.}},
\bauthor{\bsnm{{Pastorello}}, \binits{A.}},
\bauthor{\bsnm{{Smartt}}, \binits{S.J.}},
\bauthor{\bsnm{{Turatto}}, \binits{M.}},
\bauthor{\bsnm{{Yuan}}, \binits{F.}},
\bauthor{\bsnm{{Anderson}}, \binits{J.P.}},
\bauthor{\bsnm{{Bayliss}}, \binits{D.D.R.}},
\bauthor{\bsnm{{Bauer}}, \binits{F.E.}},
\bauthor{\bsnm{{Chen}}, \binits{T.-W.}},
\bauthor{\bsnm{{F{\"o}rster Bur{\'o}n}}, \binits{F.}},
\bauthor{\bsnm{{Gal-Yam}}, \binits{A.}},
\bauthor{\bsnm{{Haislip}}, \binits{J.B.}},
\bauthor{\bsnm{{Knapic}}, \binits{C.}},
\bauthor{\bsnm{{Le Guillou}}, \binits{L.}},
\bauthor{\bsnm{{Marchi}}, \binits{S.}},
\bauthor{\bsnm{{Mazzali}}, \binits{P.}},
\bauthor{\bsnm{{Molinaro}}, \binits{M.}},
\bauthor{\bsnm{{Moore}}, \binits{J.P.}},
\bauthor{\bsnm{{Reichart}}, \binits{D.}},
\bauthor{\bsnm{{Smareglia}}, \binits{R.}},
\bauthor{\bsnm{{Smith}}, \binits{K.W.}},
\bauthor{\bsnm{{Sternberg}}, \binits{A.}},
\bauthor{\bsnm{{Sullivan}}, \binits{M.}},
\bauthor{\bsnm{{Tak{\'a}ts}}, \binits{K.}},
\bauthor{\bsnm{{Tucker}}, \binits{B.E.}},
\bauthor{\bsnm{{Valenti}}, \binits{S.}},
\bauthor{\bsnm{{Yaron}}, \binits{O.}},
\bauthor{\bsnm{{Young}}, \binits{D.R.}},
\bauthor{\bsnm{{Zhou}}, \binits{G.}}:
\bjtitle{\mnras}
\bvolume{433},
\bfpage{1312}
(\byear{2013}).
\arxivurl{1303.3453}.
doi:\doiurl{10.1093/mnras/stt813}
\end{barticle}
\endbibitem

\bibitem[\protect\citeauthoryear{{Green} et~al.}{2010}]{green10}
\begin{barticle}
\bauthor{\bsnm{{Green}}, \binits{A.W.}},
\bauthor{\bsnm{{Glazebrook}}, \binits{K.}},
\bauthor{\bsnm{{McGregor}}, \binits{P.J.}},
\bauthor{\bsnm{{Abraham}}, \binits{R.G.}},
\bauthor{\bsnm{{Poole}}, \binits{G.B.}},
\bauthor{\bsnm{{Damjanov}}, \binits{I.}},
\bauthor{\bsnm{{McCarthy}}, \binits{P.J.}},
\bauthor{\bsnm{{Colless}}, \binits{M.}},
\bauthor{\bsnm{{Sharp}}, \binits{R.G.}}:
\bjtitle{\nat}
\bvolume{467},
\bfpage{684}
(\byear{2010}).
\arxivurl{1010.1262}.
doi:\doiurl{10.1038/nature09452}
\end{barticle}
\endbibitem

\bibitem[\protect\citeauthoryear{{Inserra} et~al.}{2013}]{inserra12ca}
\begin{botherref}
\oauthor{\bsnm{{Inserra}}, \binits{C.}},
\oauthor{\bsnm{{Smartt}}, \binits{S.J.}},
\oauthor{\bsnm{{Scalzo}}, \binits{R.}},
\oauthor{\bsnm{{Fraser}}, \binits{M.}},
\oauthor{\bsnm{{Pastorello}}, \binits{A.}},
\oauthor{\bsnm{{Childress}}, \binits{M.}},
\oauthor{\bsnm{{Pignata}}, \binits{G.}},
\oauthor{\bsnm{{Jerkstrand}}, \binits{A.}},
\oauthor{\bsnm{{Kotak}}, \binits{R.}},
\oauthor{\bsnm{{Benetti}}, \binits{S.}},
\oauthor{\bsnm{{Della Valle}}, \binits{M.}},
\oauthor{\bsnm{{Gal-Yam}}, \binits{A.}},
\oauthor{\bsnm{{Mazzali}}, \binits{P.}},
\oauthor{\bsnm{{Smith}}, \binits{K.}},
\oauthor{\bsnm{{Sullivan}}, \binits{M.}},
\oauthor{\bsnm{{Valenti}}, \binits{S.}},
\oauthor{\bsnm{{Yaron}}, \binits{O.}},
\oauthor{\bsnm{{Young}}, \binits{D.}}:
ArXiv e-prints
(2013).
\arxivurl{1307.1791}
\end{botherref}
\endbibitem

\bibitem[\protect\citeauthoryear{{Markwardt}}{2009}]{markwardt09}
\begin{bchapter}
\bauthor{\bsnm{{Markwardt}}, \binits{C.B.}}:
In: \beditor{\bsnm{{Bohlender}}, \binits{D.A.}},
\beditor{\bsnm{{Durand}}, \binits{D.}},
\beditor{\bsnm{{Dowler}}, \binits{P.}} (eds.)
\bbtitle{Astronomical Data Analysis Software and Systems XVIII}.
\bsertitle{Astronomical Society of the Pacific Conference Series},
vol. \bseriesno{411},
p. \bfpage{251}
(\byear{2009}).
\arxivurl{0902.2850}
\end{bchapter}
\endbibitem

\bibitem[\protect\citeauthoryear{{Maund} et~al.}{2013}]{maund12ec}
\begin{barticle}
\bauthor{\bsnm{{Maund}}, \binits{J.R.}},
\bauthor{\bsnm{{Fraser}}, \binits{M.}},
\bauthor{\bsnm{{Smartt}}, \binits{S.J.}},
\bauthor{\bsnm{{Botticella}}, \binits{M.T.}},
\bauthor{\bsnm{{Barbarino}}, \binits{C.}},
\bauthor{\bsnm{{Childress}}, \binits{M.}},
\bauthor{\bsnm{{Gal-Yam}}, \binits{A.}},
\bauthor{\bsnm{{Inserra}}, \binits{C.}},
\bauthor{\bsnm{{Pignata}}, \binits{G.}},
\bauthor{\bsnm{{Reichart}}, \binits{D.}},
\bauthor{\bsnm{{Schmidt}}, \binits{B.}},
\bauthor{\bsnm{{Sollerman}}, \binits{J.}},
\bauthor{\bsnm{{Taddia}}, \binits{F.}},
\bauthor{\bsnm{{Tomasella}}, \binits{L.}},
\bauthor{\bsnm{{Valenti}}, \binits{S.}},
\bauthor{\bsnm{{Yaron}}, \binits{O.}}:
\bjtitle{\mnras}
\bvolume{431},
\bfpage{102}
(\byear{2013}).
\arxivurl{1302.0170}.
doi:\doiurl{10.1093/mnrasl/slt017}
\end{barticle}
\endbibitem

\bibitem[\protect\citeauthoryear{{McGregor} et~al.}{2003}]{mcgregor03}
\begin{bchapter}
\bauthor{\bsnm{{McGregor}}, \binits{P.J.}},
\bauthor{\bsnm{{Hart}}, \binits{J.}},
\bauthor{\bsnm{{Conroy}}, \binits{P.G.}},
\bauthor{\bsnm{{Pfitzner}}, \binits{M.L.}},
\bauthor{\bsnm{{Bloxham}}, \binits{G.J.}},
\bauthor{\bsnm{{Jones}}, \binits{D.J.}},
\bauthor{\bsnm{{Downing}}, \binits{M.D.}},
\bauthor{\bsnm{{Dawson}}, \binits{M.}},
\bauthor{\bsnm{{Young}}, \binits{P.}},
\bauthor{\bsnm{{Jarnyk}}, \binits{M.}},
\bauthor{\bsnm{{Van Harmelen}}, \binits{J.}}:
In: \beditor{\bsnm{{Iye}}, \binits{M.}},
\beditor{\bsnm{{Moorwood}}, \binits{A.F.M.}} (eds.)
\bbtitle{Society of Photo-Optical Instrumentation Engineers (SPIE) Conference
  Series}.
\bsertitle{Society of Photo-Optical Instrumentation Engineers (SPIE) Conference
  Series},
vol. \bseriesno{4841},
p. \bfpage{1581}
(\byear{2003}).
doi:\doiurl{10.1117/12.459448}
\end{bchapter}
\endbibitem

\bibitem[\protect\citeauthoryear{{McGregor} et~al.}{2012}]{mcgregor12}
\begin{bchapter}
\bauthor{\bsnm{{McGregor}}, \binits{P.J.}},
\bauthor{\bsnm{{Bloxham}}, \binits{G.J.}},
\bauthor{\bsnm{{Boz}}, \binits{R.}},
\bauthor{\bsnm{{Davies}}, \binits{J.}},
\bauthor{\bsnm{{Doolan}}, \binits{M.}},
\bauthor{\bsnm{{Ellis}}, \binits{M.}},
\bauthor{\bsnm{{Hart}}, \binits{J.}},
\bauthor{\bsnm{{Jones}}, \binits{D.J.}},
\bauthor{\bsnm{{Luvaul}}, \binits{L.}},
\bauthor{\bsnm{{Nielsen}}, \binits{J.}},
\bauthor{\bsnm{{Parcell}}, \binits{S.}},
\bauthor{\bsnm{{Sharp}}, \binits{R.}},
\bauthor{\bsnm{{Stevanovic}}, \binits{D.}},
\bauthor{\bsnm{{Young}}, \binits{P.J.}}:
In: \bbtitle{Ground-based and Airborne Instrumentation for Astronomy IV}.
\bsertitle{Society of Photo-Optical Instrumentation Engineers (SPIE) Conference
  Series},
vol. \bseriesno{8446},
\bconfdate{September} \byear{2012}.
doi:\doiurl{10.1117/12.925259}
\end{bchapter}
\endbibitem

\bibitem[\protect\citeauthoryear{{Oke}}{1990}]{oke90}
\begin{barticle}
\bauthor{\bsnm{{Oke}}, \binits{J.B.}}:
\bjtitle{\aj}
\bvolume{99},
\bfpage{1621}
(\byear{1990}).
doi:\doiurl{10.1086/115444}
\end{barticle}
\endbibitem

\bibitem[\protect\citeauthoryear{Press et~al.}{2007}]{press07}
\begin{bbook}
\bauthor{\bsnm{Press}, \binits{W.H.}},
\bauthor{\bsnm{Teukolsky}, \binits{S.A.}},
\bauthor{\bsnm{Vetterling}, \binits{W.T.}},
\bauthor{\bsnm{Flannery}, \binits{B.P.}}:
\bbtitle{Numerical Recipes 3rd Edition: The Art of Scientific Computing},
\bedition{3}rd edn.
\bpublisher{Cambridge University Press}, \blocation{???}
(\byear{2007})
\end{bbook}
\endbibitem

\bibitem[\protect\citeauthoryear{{Rich} et~al.}{2010}]{rich10}
\begin{barticle}
\bauthor{\bsnm{{Rich}}, \binits{J.A.}},
\bauthor{\bsnm{{Dopita}}, \binits{M.A.}},
\bauthor{\bsnm{{Kewley}}, \binits{L.J.}},
\bauthor{\bsnm{{Rupke}}, \binits{D.S.N.}}:
\bjtitle{\apj}
\bvolume{721},
\bfpage{505}
(\byear{2010}).
\arxivurl{1007.3495}.
doi:\doiurl{10.1088/0004-637X/721/1/505}
\end{barticle}
\endbibitem

\bibitem[\protect\citeauthoryear{{Savitzky} and {Golay}}{1964}]{savitzky64}
\begin{barticle}
\bauthor{\bsnm{{Savitzky}}, \binits{A.}},
\bauthor{\bsnm{{Golay}}, \binits{M.J.E.}}:
\bjtitle{Analytical Chemistry}
\bvolume{36},
\bfpage{1627}
(\byear{1964})
\end{barticle}
\endbibitem

\bibitem[\protect\citeauthoryear{{Spergel} et~al.}{2013}]{spergel13}
\begin{botherref}
\oauthor{\bsnm{{Spergel}}, \binits{D.}},
\oauthor{\bsnm{{Gehrels}}, \binits{N.}},
\oauthor{\bsnm{{Breckinridge}}, \binits{J.}},
\oauthor{\bsnm{{Donahue}}, \binits{M.}},
\oauthor{\bsnm{{Dressler}}, \binits{A.}},
\oauthor{\bsnm{{Gaudi}}, \binits{B.S.}},
\oauthor{\bsnm{{Greene}}, \binits{T.}},
\oauthor{\bsnm{{Guyon}}, \binits{O.}},
\oauthor{\bsnm{{Hirata}}, \binits{C.}},
\oauthor{\bsnm{{Kalirai}}, \binits{J.}},
\oauthor{\bsnm{{Kasdin}}, \binits{N.J.}},
\oauthor{\bsnm{{Moos}}, \binits{W.}},
\oauthor{\bsnm{{Perlmutter}}, \binits{S.}},
\oauthor{\bsnm{{Postman}}, \binits{M.}},
\oauthor{\bsnm{{Rauscher}}, \binits{B.}},
\oauthor{\bsnm{{Rhodes}}, \binits{J.}},
\oauthor{\bsnm{{Wang}}, \binits{Y.}},
\oauthor{\bsnm{{Weinberg}}, \binits{D.}},
\oauthor{\bsnm{{Centrella}}, \binits{J.}},
\oauthor{\bsnm{{Traub}}, \binits{W.}},
\oauthor{\bsnm{{Baltay}}, \binits{C.}},
\oauthor{\bsnm{{Colbert}}, \binits{J.}},
\oauthor{\bsnm{{Bennett}}, \binits{D.}},
\oauthor{\bsnm{{Kiessling}}, \binits{A.}},
\oauthor{\bsnm{{Macintosh}}, \binits{B.}},
\oauthor{\bsnm{{Merten}}, \binits{J.}},
\oauthor{\bsnm{{Mortonson}}, \binits{M.}},
\oauthor{\bsnm{{Penny}}, \binits{M.}},
\oauthor{\bsnm{{Rozo}}, \binits{E.}},
\oauthor{\bsnm{{Savransky}}, \binits{D.}},
\oauthor{\bsnm{{Stapelfeldt}}, \binits{K.}},
\oauthor{\bsnm{{Zu}}, \binits{Y.}},
\oauthor{\bsnm{{Baker}}, \binits{C.}},
\oauthor{\bsnm{{Cheng}}, \binits{E.}},
\oauthor{\bsnm{{Content}}, \binits{D.}},
\oauthor{\bsnm{{Dooley}}, \binits{J.}},
\oauthor{\bsnm{{Foote}}, \binits{M.}},
\oauthor{\bsnm{{Goullioud}}, \binits{R.}},
\oauthor{\bsnm{{Grady}}, \binits{K.}},
\oauthor{\bsnm{{Jackson}}, \binits{C.}},
\oauthor{\bsnm{{Kruk}}, \binits{J.}},
\oauthor{\bsnm{{Levine}}, \binits{M.}},
\oauthor{\bsnm{{Melton}}, \binits{M.}},
\oauthor{\bsnm{{Peddie}}, \binits{C.}},
\oauthor{\bsnm{{Ruffa}}, \binits{J.}},
\oauthor{\bsnm{{Shaklan}}, \binits{S.}}:
ArXiv e-prints
(2013).
\arxivurl{1305.5422}
\end{botherref}
\endbibitem

\bibitem[\protect\citeauthoryear{{van Dokkum}}{2001}]{vandokkum01}
\begin{barticle}
\bauthor{\bsnm{{van Dokkum}}, \binits{P.G.}}:
\bjtitle{\pasp}
\bvolume{113},
\bfpage{1420}
(\byear{2001}).
\arxivurl{arXiv:astro-ph/0108003}.
doi:\doiurl{10.1086/323894}
\end{barticle}
\endbibitem

\bibitem[\protect\citeauthoryear{{Vogt} et~al.}{2013}]{vogt13}
\begin{barticle}
\bauthor{\bsnm{{Vogt}}, \binits{F.P.A.}},
\bauthor{\bsnm{{Dopita}}, \binits{M.A.}},
\bauthor{\bsnm{{Kewley}}, \binits{L.J.}}:
\bjtitle{\apj}
\bvolume{768},
\bfpage{151}
(\byear{2013}).
\arxivurl{1303.0290}.
doi:\doiurl{10.1088/0004-637X/768/2/151}
\end{barticle}
\endbibitem

\end{thebibliography}

%

\end{document}